\documentclass[journal]{IEEEtran}
\ifCLASSINFOpdf
\else
\fi

\usepackage{color}
\usepackage{url}
\usepackage{graphicx}
\usepackage[cmex10]{amsmath}
\usepackage{subfloat}
\usepackage{subfig}
\usepackage{overpic}
\usepackage{amsmath}
\usepackage[normalem]{ulem}

\begin{document}
%
\title{A Billboard-based Free-viewpoint Video Synthesizing Algorithm for Sports Scenes}
%
%
%

\author{Jun~Chen,~
        Ryosuke~Watanabe,~
        Keisuke~Nonaka,~
        Tomoaki~Konno,~
        Hiroshi~Sankoh,~
        and~Sei~Naito~
\thanks{
J. Chen, R. Watanabe, K. Nonaka, T. Konno, H. Sankoh, S. Naito
are with Ultra-realistic Communication Group, KDDI Research, Inc., Fujimino, Japan (corresponding author (J. Chen) Tel: +81-70-3825-9914; e-mail: ju-chen@kddi-research.jp).}}


%
%

\markboth{Journal of \LaTeX\ Class Files,~Vol.~14, No.~8, August~2015}%
{Shell \MakeLowercase{\textit{et al.}}: Bare Demo of IEEEtran.cls for IEEE Journals}
%



\maketitle

\begin{abstract}

We present a billboard-based free-viewpoint video synthesizing algorithm for sports scenes that can robustly reconstruct and render a high-fidelity billboard model for each object, including an occluded one, in each camera.
Its contributions are
(1) applicable to a challenging shooting situation where a high precision 3D model cannot be built because only a small number of cameras, featuring wide-baseline are available;
(2) capable of reproducing the appearance of occlusions, which is one of the most significant issues for billboard-based approaches due to the ineffective detection of overlaps.
To achieve these goals above, the proposed method does not attempt to find a high-quality 3D model but utilizes a raw 3D model that is obtained directly from space carving.
Although the model is insufficiently accurate for producing an impressive visual effect, precise object segmentation and occlusions detection can be performed by back-projection onto each camera plane.
The billboard model of each object in each camera is rendered according to whether it is occluded or not, and its location in the virtual stadium is determined by considering the barycenter of its 3D model.
We synthesized free-viewpoint videos of two soccer sequences recorded by five cameras, using the proposed and state-of-the-art methods to demonstrate the effectiveness of the proposed method.
 
\end{abstract}


\section{Introduction}

\IEEEPARstart{F}{ree-viewpoint} video synthesis is an active research field in computer vision, aimed at providing a beyond-3D experience, in which audiences can view virtual media from any preferred angle and position.
In a free-viewpoint video system, the virtual viewpoint can be interactively selected to see a part of the field from angles where a camera cannot be mounted. 
Moreover, the viewpoint can be moved around the stadium to allow audiences to have a walk-through or fly-through experience \cite{tanimoto2006overview, smolic20113d, kanade1997virtualized, matsuyama2002generation}. 

{The primary way to produce such a visual effect is to equip the observed scene with a synchronized camera-network {\cite{Sankoh2018Acmmm, nonaka2018, chen2019fast}}.
A free-viewpoint video is then created by using multi-view geometry techniques, such as 3D reconstruction or view-dependent representation.
The 3D model representation, by means of a 3D mesh or point cloud \cite{theobalt2004model, kilner2007dual, liu2010point, collet2015high}, provides full freedom of virtual view and continuous appearance changes for objects. 
Therefore, this representation is close to the original concept of a free-viewpoint video.
An example of this technology is the "Intel True View" for Super Bowl LIII \cite{InterTrueView} that enables immersive viewing experiences by transforming video data captured from $38$ $5$K ultra-high-definition cameras into a 3D video.
This technology achieves impressive results.
However, a camera-network with many well-calibrated cameras is required to obtain a precise model.
This makes these methods difficult to deploy cost-effectively.
{Moreover, the heavy computational process of rendering leads to a non-real-time video display, especially for portable devices like smartphones.}
{The view-dependent representation techniques \cite{germann2010articulated, inamoto2007virtual, Sabirin2018Toward, Sankoh2018Acmmm} do not provide a consistent solution for all input cameras, but compute a separate reconstruction for each viewpoint.}
In general, these techniques do not require a large number of cameras.
As reported in \cite{germann2012novel}, a novel view can be synthesized employing only two cameras by using sparse point correspondences and a coarse-to-fine reconstruction method.
The requirement for numerous physical devices was relaxed. 
But at the same time, this introduces new challenges.
{The biggest challenge in these methods is the detection and rendering of ``occlusion", which is the overlap of multiple objects in a camera view.}

With the convergence of technologies from computer vision and deep learning \cite{he2017mask, wei2016convolutional}, an alternative way to create a free-viewpoint video is to convert a single camera signal into a proper 3D representation \cite{rematas2018soccer, mustafa2017semantically}.
The new way makes a creation easily controllable, flexible, convenient, and cheap.
As noted in \cite{rematas2018soccer}, it uses a CNN to estimate a player body depth map to reconstruct a soccer game from just a single YouTube video.
Despite their generality, however, there are numerous challenges in this setup due to several factors.
First, it cannot reproduce an appropriate appearance over the entire range of virtual views due to the limited information.
For example, the surface texture of an opposite side, beyond the camera's sight, is unlikely to produce a satisfactory visual effect. 
The detection and treatment of occlusions caused by overlaps of multiple objects in a single camera view remain to be solved.
Also, errors in occlusion detection lead to inaccurate depth estimation.

In this paper, we focus on a multi-camera setup to provide an immersive free-viewpoint video for a sports scene, such as soccer or rugby, that involves a large field.
Its goal is to resolve the conflicting creation of a high-fidelity free-viewpoint video with the requirement for many cameras.
To be specific, we proposed an algorithm to robustly reconstruct an accurate billboard model for each object, including occluded ones, in each camera. 
It can be applied to challenging shooting conditions where only a few cameras featuring wide-baseline are present.
Our key ideas are: (1) accurate depth estimation and object segmentation are achieved by projecting labelled 3D models, obtained from shape-from-silhouette without optimization, onto each camera plane;
(2) the occlusion of each object is detected using the acquired 2D segmentation map without the involvement of parameters and robust against self-occlusion;
(3) a reasonable 3D coordinate of each billboard model in a virtual stadium is calculated according to the barycenter of the raw 3D model to provide a stereovision effect.

We present the synthesized results of two soccer contents that were recorded by {five cameras}.
Our results can be viewed on a PC, smartphone, smartglass, or head-mounted display, enabling free-viewpoint navigation to any virtual viewpoint.
Comparative results are also provided to show the effectiveness of the proposed method in terms of the naturalness of the surface appearance in the synthesized billboard models.

\section{Related Works}

\subsection{Free-viewpoint Video Creation from Multiple Views}
\label{sec:related_work_A}

\subsubsection{3D model Representation}

The visual hull \cite{laurentini1994visual, cheung2000real, ladikos2008efficient} is a 3D reconstruction technique that approximates the shape of observed objects from a set of calibrated silhouettes.
{It usually discretizes a pre-defined 3D volume into voxels and tests whether a voxel is available or not by determining whether it falls inside or outside the silhouettes.}
Coupled with the marching cubes algorithm \cite{lorensen1987marching, newman2006survey}, the discrete voxel representation can be converted into a triangle mesh form.
Some approaches focus on the direct calculation of a mesh representation by analyzing the geometric relation between silhouettes and a visual hull surface based on the assumption of local smoothness or point-plane duality \cite{liang2005complex, franco2009efficient}.
{Visual hull approaches suffer from two main limitations. 
First, many calibrated cameras need to be placed in a $360$-degree circle to obtain a relatively precise model. 
Second, it gives the maximal volume consistent with objects' silhouettes, failing to reconstruct concavities.}
More generally, visual hull approaches serve as initialization for more elaborate 3D reconstruction.
The photo-hull \cite{slabaugh2002image, fan2010photo, slabaugh2005image} approximates the maximum boundaries of objects using photo-consistency of a set of calibrated images.
It eliminates the process of silhouette extraction but introduces more restrictions, such as highly precise camera calibration, sufficient texture, and diffuse surface reflectance.
As noted in \cite{furukawa2009carved, 8237598, starck2007surface, robertini2016model, loper2014mosh}, advanced approaches combine photo-consistency, silhouette-consistency, sparse feature correspondence, and more, to solve the problem of high-quality reconstruction.
However, it takes time to process parameter-tuning to balance the constraints.

\subsubsection{View-dependent Representation}

View-dependent representations can be classified into view interpolation and billboard-based methods by their different procedures. 
View interpolation \cite{inamoto2007virtual, li2009virtual, 7997902} utilizes the projective geometry between neighboring cameras to synthesize a view without explicit reconstruction of a 3D model.
It has the advantage of avoiding the processes of camera calibration and 3D model estimation. 
{However, the quality of a synthesized view is restricted by the accuracy of the correspondences among cameras, which means that the optimal baseline is constrained in a relatively narrow range.
An interpolation method \cite{lipski2014correspondence} that renders a scene using both pixel correspondence and a depth map was reported to improve the visual effect. 
Nevertheless, it still suffers from a narrow baseline.}
Billboard-based methods {\cite{germann2010articulated, Apractical2015, 8081459, nonaka2018optimal}} construct a single planar billboard for each object in each camera. 
The billboards rotate around individual points of the virtual stadium as the viewpoint moves, providing walk-through and fly-through experiences. 
These methods cannot reproduce continuous changes in the appearance of an object, but the representation can easily be reconstructed.
{
{Our previous work \cite{Sankoh2018Acmmm} overcomes the problem of occlusion by utilizing conservative 3D models to segment objects.
Its underlying assumption is that the back-projection area of a conservative 3D model in a camera is always larger than the input silhouette.
It outperforms conventional methods in terms of robustness on camera setup and naturalness of texture.
However, we find that the reconstruction of rough 3D models increases noise and degrades the final visual effect.}

\subsection{Free-viewpoint Video Creation from a Single View}

Creating a free-viewpoint video from a single camera (generally a moving camera) is a delicate task, which involves automatic camera calibration, semantic segmentation, and monocular depth estimation.
The calibration methods {\cite{yao2017fast, yao2016automatic, farin2003robust}} are generally composed of three processes, including field line extraction, cross point calculation, and field model matching.
{With an assumption of small movement between consecutive frames, \cite{yao2016robust} calibrates the first frame using conventional methods and propagates the parameters of the current frame from previous frames by estimating the homographic matrix.}
Semantic segmentation \cite{long2015fully, chen2017rethinking, he2017mask} is a pixel-level dense prediction task that labels each pixel of an image with a corresponding class of what is being represented.
{In an application of free-viewpoint video creation, it works out what objects there are, and where are they in an image, to the information needed for further processing.}
Estimating depth is a crucial step in scene reconstruction. 
Unlike the estimation approach in multiple views that can use the correspondences among cameras, monocular depth estimation \cite{fu2018deep, xu2018monocular, bhoi2019monocular} is a technique of estimating depth from a single RGB image.
Many recent works \cite{Pavlakos_2017, Tome_2017, Zhou_2017} follow an end-to-end learning paradigm consisting of a Convolutional Network for 2D/3D body joint localization and a subsequent optimization step to regress to a 3D pose.
{The constraint on these methods is the requirement of images with 2D/3D pose ground truth for training.}
The study \cite{rematas2018soccer} presented here describes the first-ever method that can transform a monocular video of a soccer game into a free-viewpoint video by combining the techniques mentioned above.
It constructs a dataset of depth-map / image pairs from FIFA video game for the restricted soccer scenario to improve the accuracy of depth estimation.
{The approach reported in \cite{Sabirin2018Toward} can also create a free-viewpoint video from a single video.
The major deficiency of creation from a single view is that it can not reproduce any surface appearance that the camera does not observe.}

\begin{figure*}[t]
\centering
\includegraphics[width=0.99\linewidth]{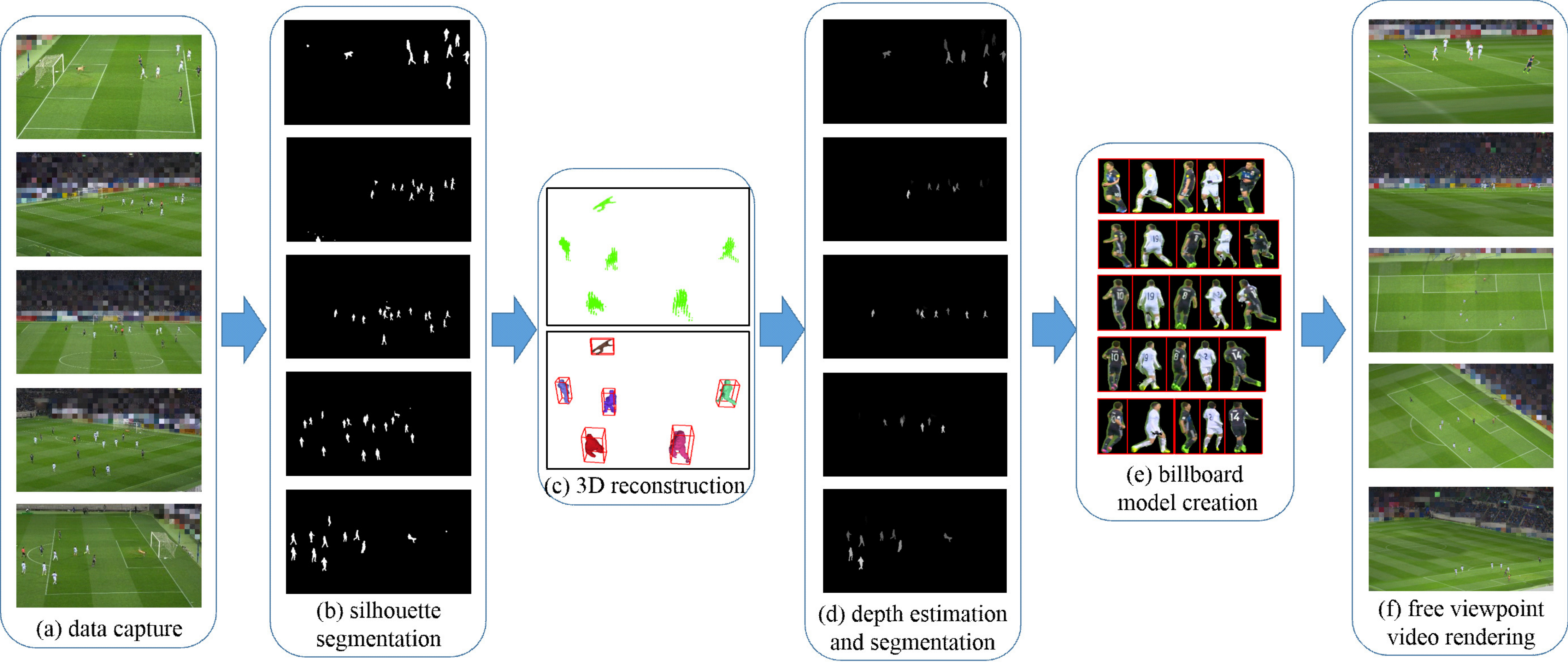}
\caption{Workflow of the proposed method.}
\label{fig:overview}
\end{figure*}%

\section{Algorithm for Free-viewpoint Video Creation}

An overview of our proposed solution is shown in Fig.~\ref{fig:overview}.
{It includes six steps: data capturing, silhouette segmentation, 3D reconstruction, depth estimation and 2D segmentation, billboard model creation, and free-viewpoint video rendering.
Processes (b)-(e) work off-line in a server-side, while the rendering is performed in real-time on the client-side according to the user's operation.}
The input data are captured using a synchronized camera network, in which the camera view is fixed during recording.
Each camera is calibrated by the method reported in \cite{yao2016automatic} to estimate the extrinsic parameters, intrinsic parameters, and lens distortion.

\subsection{Silhouette Segmentation}

{For a sports scene, it is reasonable to assume that the objects, including players and ball, are moving.
Therefore, objects can be extracted by a background subtraction method \cite{yao2015accurate} that includes three processes: global extraction, classification, and local refinement.}
In the first process, a background image is obtained by taking the average of hundreds of consecutive video frames. 
The difference in pixels between each frame and the background image is then calculated.
The pixel positions whose differences are less than a certain threshold are regarded as background, with the remaining pixels judged to be foreground. 
In the second process, we classify the shadow area into independent shadow and dependent shadow according to the shadow's luminance, shape, and size.
The independent shadows are removed here.
Finally, a refinement is conducted to remove the dependent shadow based on the assumption that the chrominance difference between objects and background is recognizable.
The threshold is adjusted dynamically according to the chrominance in each local area.

\subsection{Raw 3D Model Reconstruction}

Our method to estimate the 3D shape of observed objects from a wide-baseline camera network is to use an algorithm of shape from silhouettes. 
It discretizes a pre-defined 3D volume into voxels, projects each voxel onto all the camera image planes, and removes the voxels that fall outside the silhouettes.
The set of remaining voxels called a volumetric visual hull {\cite{laurentini1994visual}} gives a shape approximation to the observed scene.

After a volumetric visual hull is obtained, the individual objects are segmented employing a connected components labeling algorithm \cite{chen2018efficient}, and an identifier label is assigned to each object.
We extract the $0$th- and $1$st-order moment $M_{\alpha,\beta,\gamma}(\mathcal{V}_{t}) \{(\alpha,\beta,\gamma)=(0,0,0),(1,0,0),(0,1,0),(0,0,1)\}$ of each object with Eq.~\ref{equ:moment} to determine their sizes and locations with Eq.~\ref{equ:sizeAndLocation}.
\begin{eqnarray}
M_{\alpha,\beta,\gamma}(\mathcal{V}_{t})
=
\sum_{(x,y,z) \in \mathcal{V}_{t}}x^{\alpha}y^{\beta}z^{\gamma}
 {.}
\label{equ:moment}
\end{eqnarray}
\begin{equation}
\begin{aligned}
\{ N(&\mathcal{V}_{t}),X(\mathcal{V}_{t}), Y(\mathcal{V}_{t}),Z(\mathcal{V}_{t}) \} 
= \\
& \{ M_{0,0,0} (\mathcal{V}_{t}), 
\frac{M_{1,0,0}(\mathcal{V}_{t})}{M_{0,0,0}(\mathcal{V}_{t})}, 
\frac{M_{0,1,0}(\mathcal{V}_{t})}{M_{0,0,0}(\mathcal{V}_{t})}, 
\frac{M_{0,0,1}(\mathcal{V}_{t})}{M_{0,0,0}(\mathcal{V}_{t})}\}
 {.}
\label{equ:sizeAndLocation}
\end{aligned}
\end{equation}
Here, $\mathcal{V}_{t}$ expresses the $t$th object.
$\{ x,y,z \}$ denotes the 3D coordinate of an occupied voxel.
$N(\mathcal{V}_{t})$ and $\{ X(\mathcal{V}_{t}), Y(\mathcal{V}_{t}),Z(\mathcal{V}_{t}) \}$ indicate the number of voxels in $\mathcal{V}_{t}$ and its barycenter, respectively.
In the next step, we build mesh models by coupling a volumetric visual hull with a marching cubes algorithm \cite{lorensen1987marching}.

The visual hull may contain noise that comes from imperfect silhouettes.
We remove such noisy regions, taking into account the number of voxels of an object as illustrated in the following equation:
\begin{eqnarray}
\mathcal{V}_{t} =
\begin{cases} 
OFF,  & \mbox{if }  T_{min}<N(\mathcal{V}_{t})<T_{max} \\
ON, & \mbox{otherwise}
\end{cases} {.}
\label{equ:noisefilter}
\end{eqnarray}
An object is removed if its number of voxels is less than a minimum threshold $T_{min}$ or exceeds a maximum threshold $T_{max}$.
Our solution focuses on outdoor sports scenes, such as soccer or rugby match, so that it is practical to give reasonable assignments to $T_{min}$ and $T_{max}$ by considering the actual sizes of ball and athletes.
The bottom image in Fig.~\ref{fig:overview} (c) presents an example of segmentation in which a unique color is assigned to each object while the up-right rectangles illustrate the minimum bounding box of each object.

\subsection{Depth Estimation and 2D Segmentation}

\begin{figure}[t]
\centering
\footnotesize
	\begin{minipage}[b]{0.78\linewidth}
 		\centering
 		\subfloat[depth map]
		{
 	 		\begin{overpic}[width=1\textwidth]
 	 			{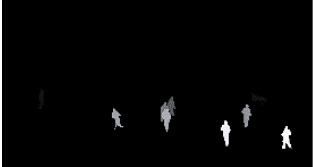}
 		\end{overpic}
 	 	}
	\end{minipage}
\vskip 1mm
	\begin{minipage}[b]{0.78\linewidth}
 		\centering
 		\subfloat[segmentation map]
		{
 			\begin{overpic}[width=1\textwidth]
 	 			{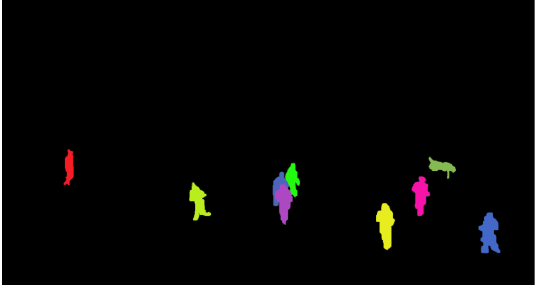}
 		\end{overpic}
 		}
	\end{minipage}			
\caption{Depth estimation and 2D segmentation.}
\label{fig:overlappingEstimation}
\end{figure}

{To estimate the depth map in a camera view, we projected the mesh models onto the camera plane to associate 2D image pixels with 3D triangles on the mesh surface.
The projection of a 3D triangle is a 2D triangle so that we defined the associations of a 3D triangle as the pixels bounded by its 2D projection.}
The depth of the $i$th pixel $d^{i}$ is assigned as the depth of the nearest corresponding triangle, as expressed in the following equation:
\begin{eqnarray}
d^{i} = \min \left\{ d^{i}_1, d^{i}_2, \cdots, d^{i}_n \right\} {.}
\label{equ:isovalueDetermination}
\end{eqnarray}
Here, $n$ indicates the number of 3D triangles that correspond with pixel $i$.
$d^{i}_j~(j=1,2,\cdots,n)$ denotes the distance from the $j$th triangle to the camera center.
Fig.~\ref{fig:overlappingEstimation} (a) presents a depth map, in which light gray coloration identifies objects as nearer to the camera center, while objects that are farther away are dark.

While estimating the depth map we also record the label of the nearest corresponding triangle, to indicate which object the pixel is associated with. 
This can be regarded as a process of segmentation in which each object is separated from the others.
Fig.~\ref{fig:overlappingEstimation} (b) demonstrates the result of segmentation, in which pixels with the same color intensity correspond to the same object.

\subsection{Billboard Model Reconstruction}

In a billboard free-viewpoint video, each object is represented as a planar image with texture, while the 3D visual effect is produced by placing the planar images in the proper position in a virtual stadium. 
In our study, we created a billboard model in the three steps described below.

\subsubsection{Individual Object Extraction}
\begin{figure}[t]
\centering		
\includegraphics[width=0.97\linewidth]{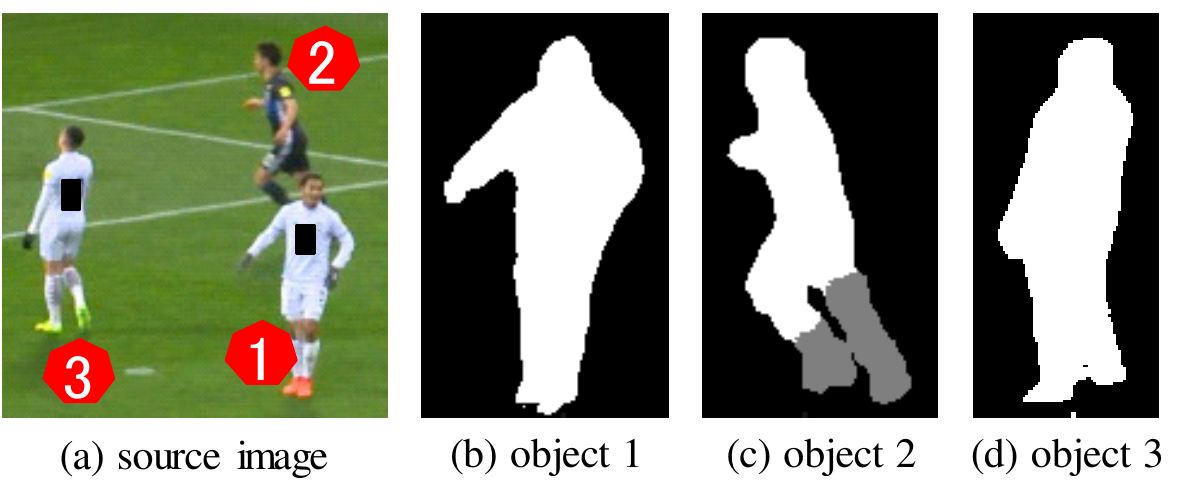}
\caption{Individual object extraction. (a) presents a cropped image where object 1 overlaps with object 2, and object 3 is isolated. (b), (c), and (d) respectively are the extracted individual objects, where the gray color indicates an occluded region. {We manually blocked the uniform number using black rectangle to avoid copyright issues. This process remains the same in the following chapters.}}
\label{fig:Individual-object-extraction}
\end{figure}
We successively project the segmented mesh models onto a specific camera plane to extract an individual 2D region for each object and determine their states, visible or not. 
The regions that map with a single object are certainly visible to the camera, while the others that are associated with two or more objects are ambiguous.
To judge the visibility of an ambiguous region, we compare the label of the projecting polygons with the label stored in the 2D segmentation map. 
It is visible when the two labels are the same. Otherwise, it is blocked by other objects. 
Fig.~\ref{fig:Individual-object-extraction} shows a demonstration in which the visible and invisible regions are respectively expressed with white and gray.
Compared with the visibility detection method using a ray-casting algorithm {\cite{Sankoh2018Acmmm} that introduces an intractable threshold, our proposed method runs without parameters and is robust against self-occlusion. 

\subsubsection{Texture Extraction}

\begin{figure}[t]
\centering
\footnotesize
	\begin{minipage}[b]{0.3029\linewidth}
 		\centering
 		\subfloat[object 1]
		{
 			\begin{overpic}[width=1\textwidth]
 	 			{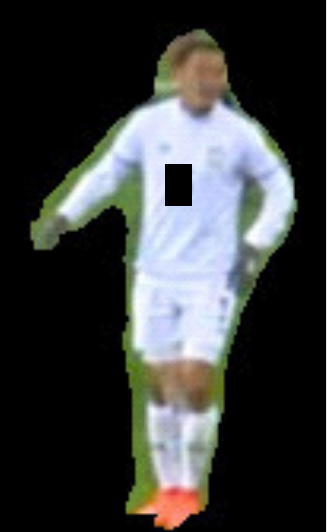}
 		\end{overpic}
 		}
	\end{minipage}	
\hskip 1mm
	\begin{minipage}[b]{0.2826\linewidth}
 		\centering
 		\subfloat[object 2]
		{
 			\begin{overpic}[width=1\textwidth]
 	 			{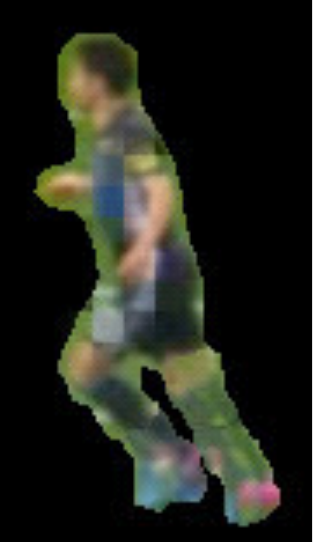}
 		\end{overpic}
 		}
	\end{minipage}	
\hskip 1mm
	\begin{minipage}[b]{0.2249\linewidth}
 		\centering
 		\subfloat[object 3]
		{
 			\begin{overpic}[width=1\textwidth]
 	 			{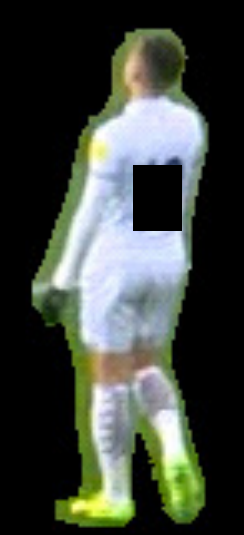}
 		\end{overpic}
 		}
	\end{minipage}			
\caption{{Texture extraction.}}
\label{fig:rendering_quality}
\vskip -2mm
\end{figure}
For the visible pixels in an individual object region, surface textures can be reproduced directly by extracting the color of the same pixels from the input image.
The invisible pixels are rendered from the neighboring cameras by coupling with the depth map and corresponding polygons.
Fig.~\ref{fig:rendering_quality} presents the rendering result of the objects in Fig.~\ref{fig:Individual-object-extraction}.
In the case of objects 1 and 3, our method produces a good appearance because their textures come from the facing camera without a blending process.
Concerning the object 2 that is partially occluded, it introduces small but acceptable visual artifacts.

\subsubsection{Location Determination}

\begin{figure}[t]
\centering
\footnotesize
	\begin{minipage}[b]{0.30\linewidth}
 		\centering
 		\subfloat[mesh model]
		{
 			\begin{overpic}[width=1\textwidth]
 	 			{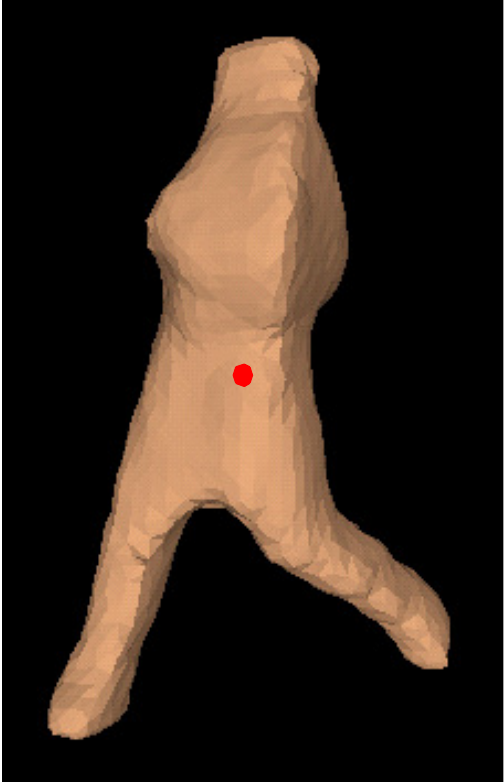}
 		\end{overpic}
 		}
	\end{minipage}	
\hskip 3mm
	\begin{minipage}[b]{0.30\linewidth}
 		\centering
 		\subfloat[billboard model]
		{
 			\begin{overpic}[width=1\textwidth]
 	 			{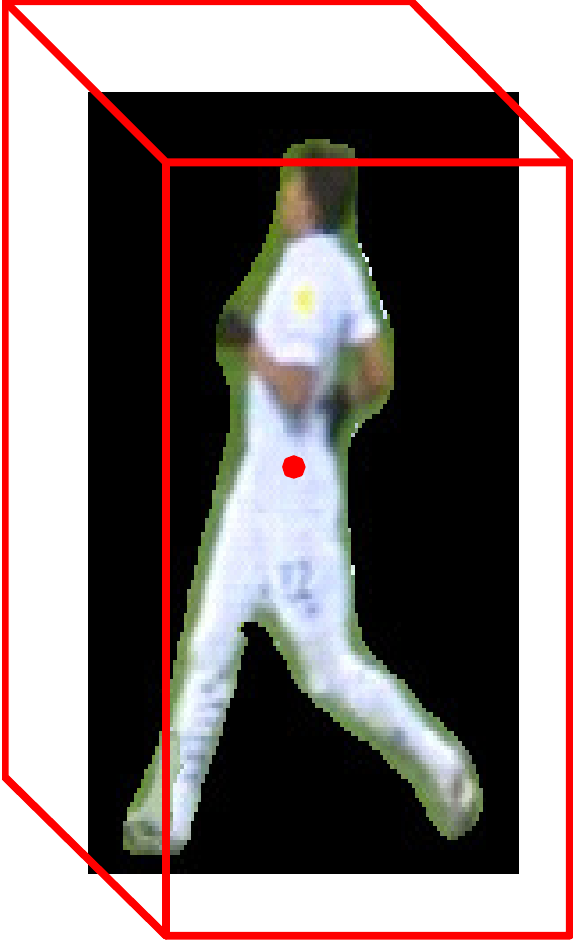}
 		\end{overpic}
 		}
	\end{minipage}			
\caption{Location determination.}
\label{fig:Location-Determination}
\end{figure} 
To accurately locate billboard models on the ground, we calculate the 2D barycentre of each object region and associate it with the 3D barycentre of its mesh model, as shown in Fig.~\ref{fig:Location-Determination}. 
The red marks respectively present the 2D and 3D barycentre while the red rectangle indicates the 3D area occupied by the mesh model.

\subsection{Free-viewpoint Video Rendering}

\begin{figure}[t]
\centering
\includegraphics[width=0.95\linewidth]{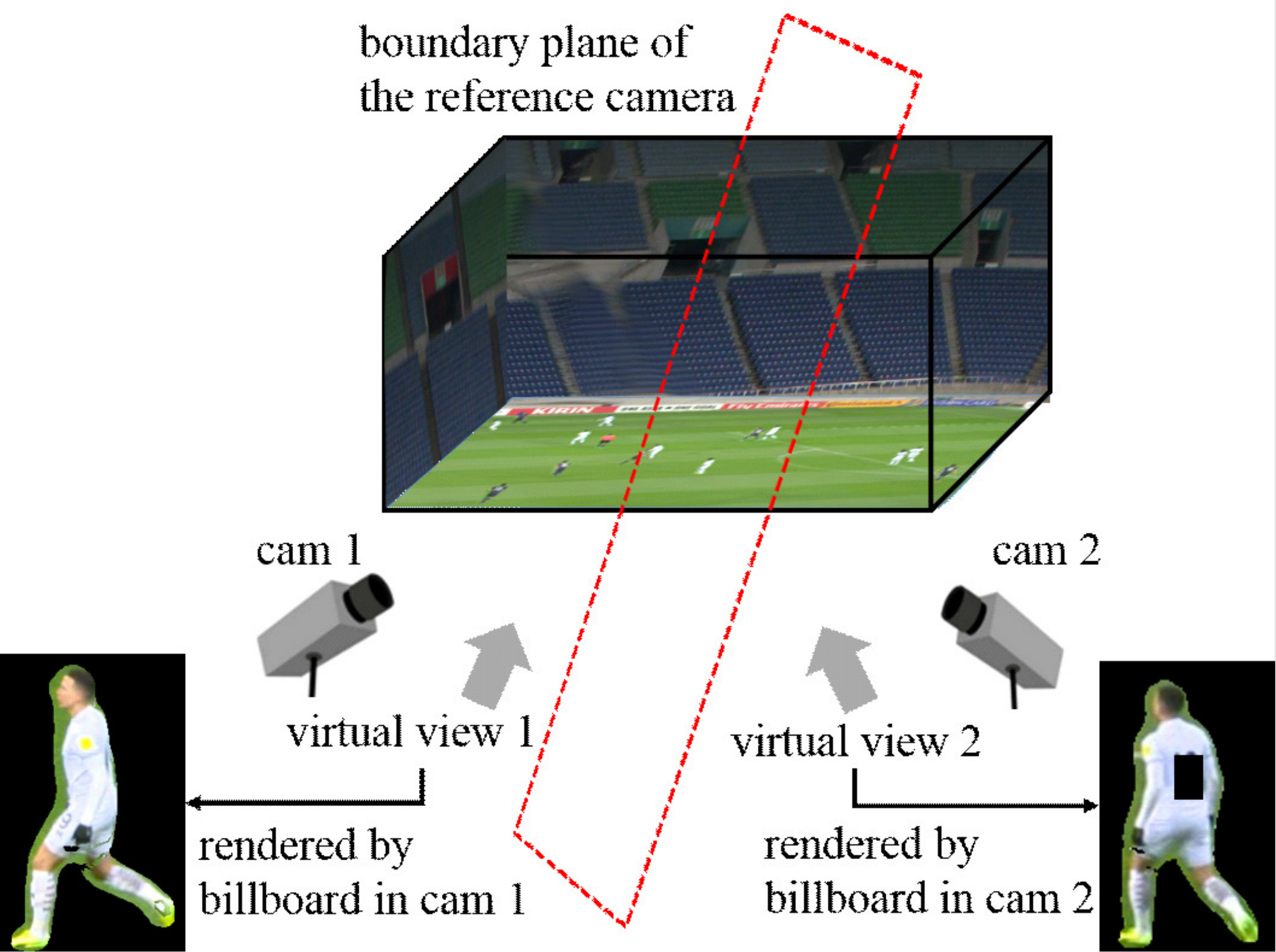}
\caption{The selection of a reference camera.}
\label{fig:fvv_rendering}
\end{figure}%

The free-viewpoint video is rendered on the client-side, where the 3D coordinate and direction of a virtual viewpoint can be obtained from the user's operation.
We identify the reference camera for rendering as the nearest camera by calculating the Euler distance between a virtual viewpoint and each recording camera.
Fig.~\ref{fig:fvv_rendering} shows an example of the selection of a reference camera.
The first camera is nearer to virtual view 1 than the second camera, so the billboards in camera 1 render its virtual image, and vice versa.
In the rendering process, billboards of a reference camera are placed in a virtual stadium and rotated according to the user-selected viewpoint.

\section{Experimental Results}

To demonstrate the performance of our method, we compare it to the following methods:
\begin{itemize}
\item[-] RB {\cite{Sankoh2018Acmmm}} as the more recent representative of the billboard-based free-viewpoint video production approach, which extracts object regions in each camera by reconstructing a rough 3D model.
\item[-] FFVV {\cite{chen2019fast}} as a more recent and fast representative of a full model free-viewpoint video generation method, which can produce a free-viewpoint video in real-time. 
\item[-] CVH {\cite{kilner2007dual}} as a conventional full model production method.
\end{itemize}

We applied the proposed and comparison methods to two types of soccer contents to validate their usability under different shooting conditions.
The vision of the cameras of the first content focuses on half of a pitch while the observation area of the second content targets the penalty area.
Both of the contents were captured with five synchronized cameras.
The resolution of each camera was $3840 \times 2160$, and the frame rate was $30$~fps.
Fig.~\ref{fig:camera-configuration} shows the camera configurations for the two contents, in which black and red symbols respectively show the position of recording cameras and virtual cameras.

\begin{figure}[t]
\centering
\footnotesize
	\begin{minipage}[b]{0.52\linewidth}
 		\centering
 		\subfloat[the first content]
		{
 			\begin{overpic}[width=1\textwidth]
 	 			{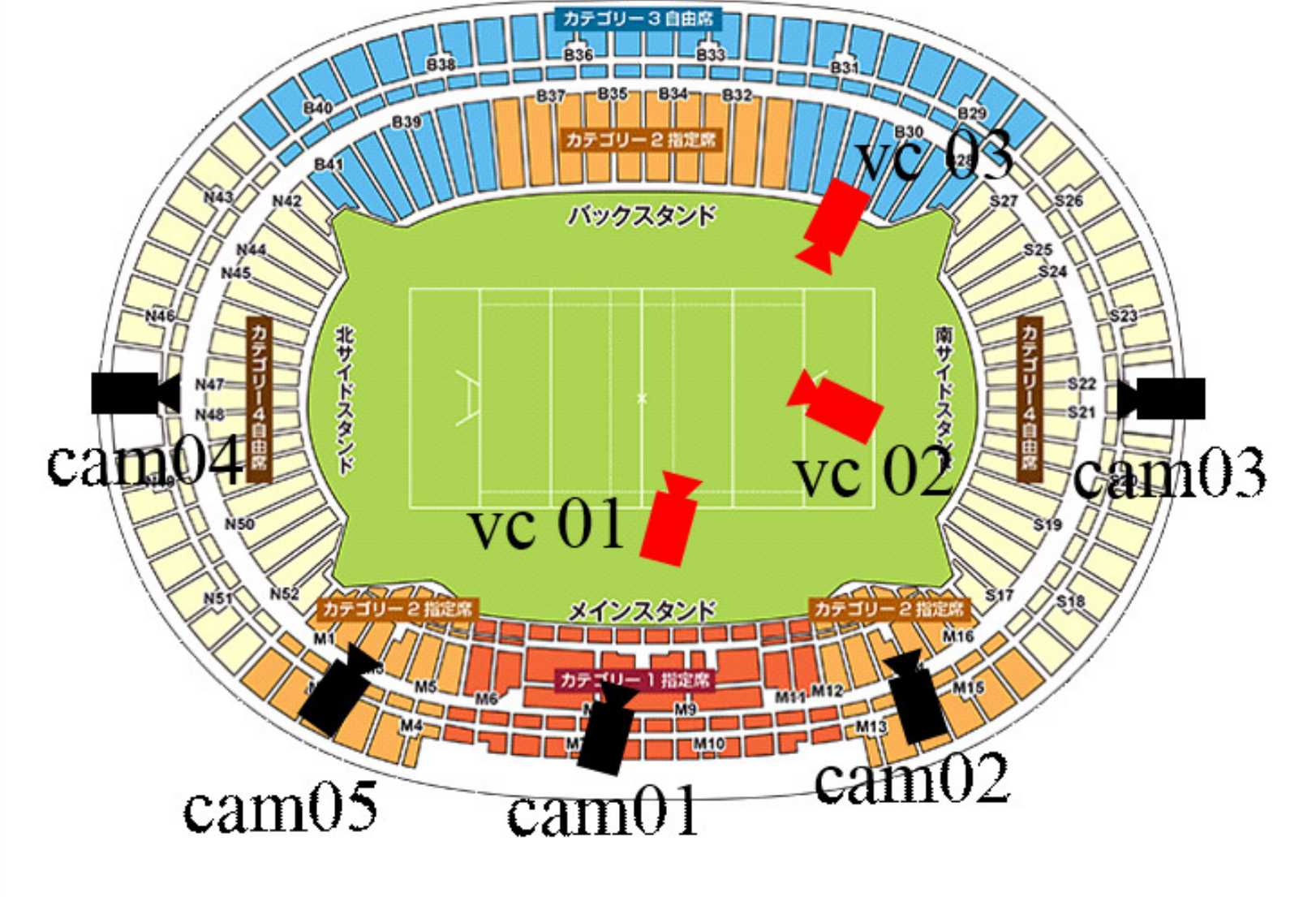}
 		\end{overpic}
 		}
	\end{minipage}	
	\begin{minipage}[b]{0.47\linewidth}
 		\centering
 		\subfloat[the second content]
		{
 			\begin{overpic}[width=1\textwidth]
 	 			{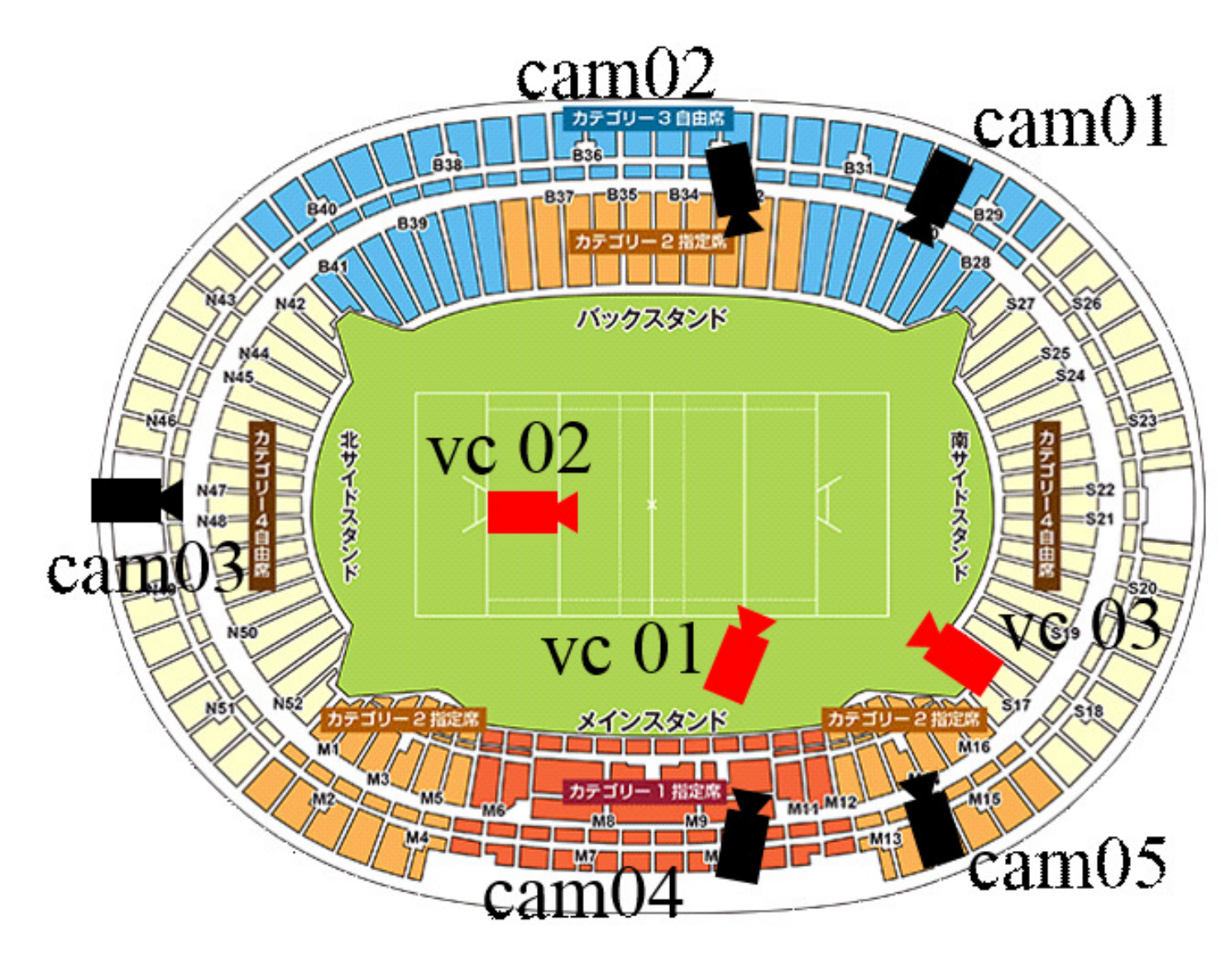}
 		\end{overpic}
 		}
	\end{minipage}			
\caption{Camera configuration.}
\label{fig:camera-configuration}
\vskip -2mm
\end{figure} 

\begin{figure*}[pt]
\centering
\footnotesize
	\begin{minipage}[b]{0.95\linewidth}
 		\centering
 		\subfloat[Cropped input images. {We manually blurred the commercial billboards to avoid copyright issues. This process remains the same for the other experiments.}]
		{
 			\begin{overpic}[width=1\textwidth]
 	 			{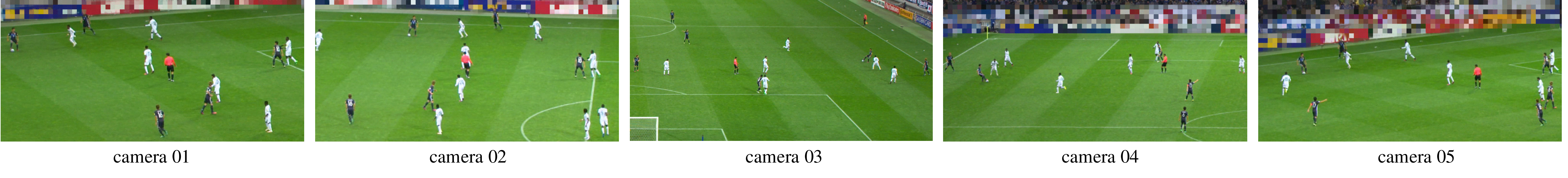}
 		\end{overpic}
 		}
	\end{minipage}	
\vskip 2mm
	\begin{minipage}[b]{0.95\linewidth}
 		\centering
 		\subfloat[Synthesized free-viewpoint video viewing from three virtual viewpoints]
		{
 			\begin{overpic}[width=1\textwidth]
 	 			{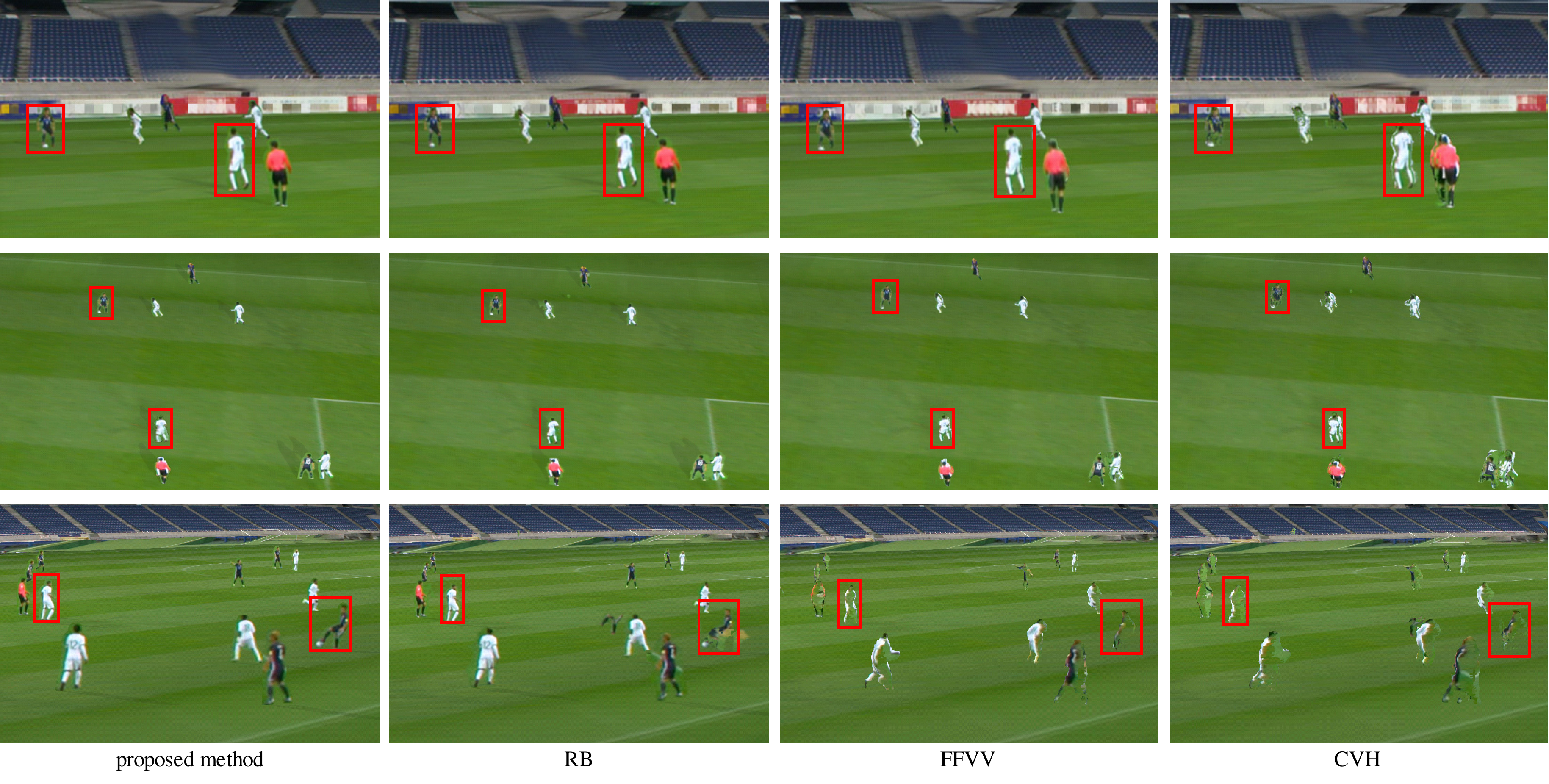}
 		\end{overpic}
 		}
	\end{minipage}	
\vskip 2mm
	\begin{minipage}[b]{0.379\linewidth}
 		\centering
 		\subfloat[Close up view of a selected player]
		{
 			\begin{overpic}[width=1\textwidth]
 	 			{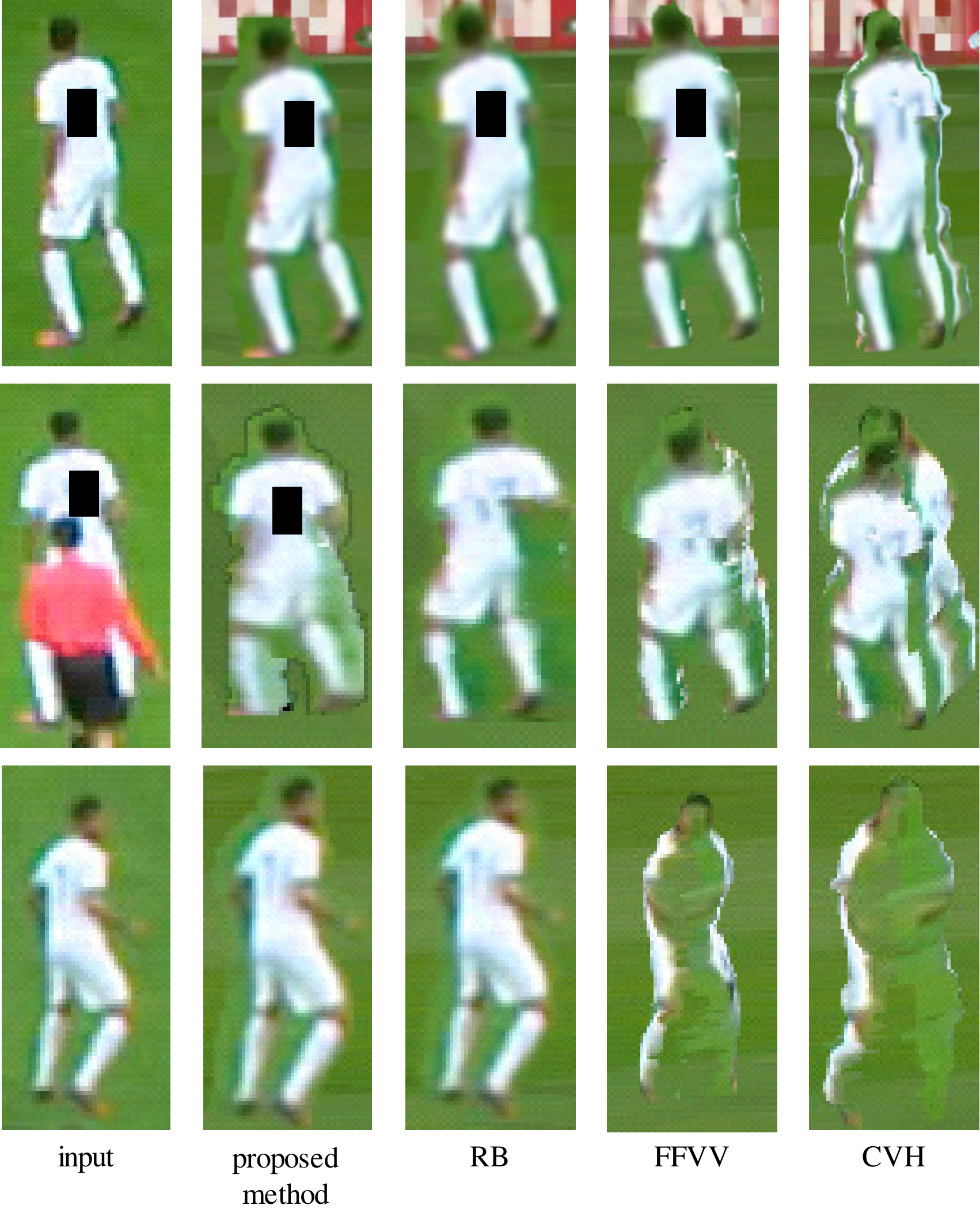}
 		\end{overpic}
 		}
	\end{minipage}	
\hskip 3mm
	\begin{minipage}[b]{0.553\linewidth}
 		\centering
 		\subfloat[Close up view of another selected player]
		{
 			\begin{overpic}[width=1\textwidth]
 	 			{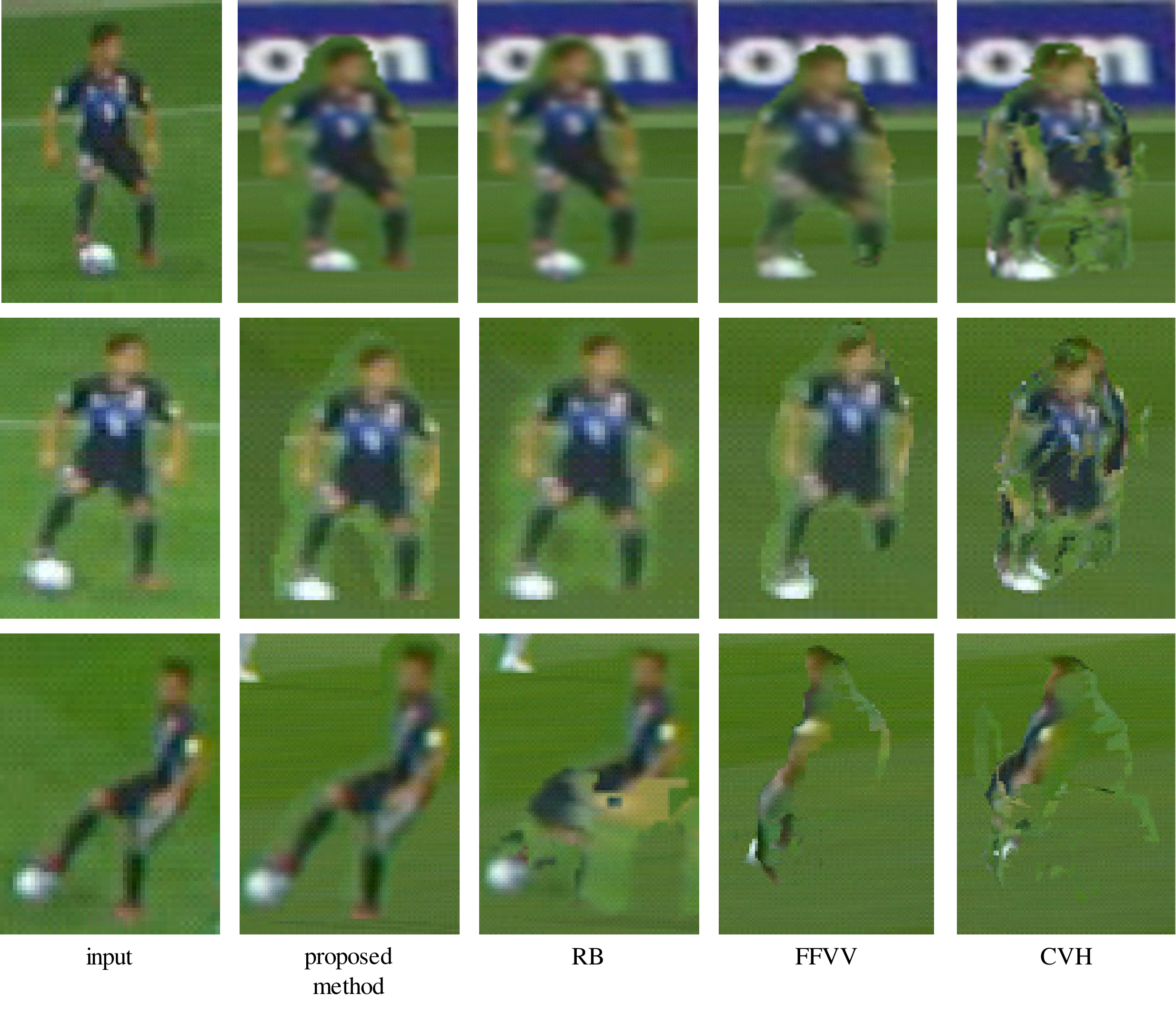}
 		\end{overpic}
 		}
	\end{minipage}	
\caption{{Free-viewpoint video of the first content}.}
\label{fig:ffv-saudi}
\end{figure*} 

\begin{figure}[pt]
\centering
\footnotesize
	\begin{minipage}[b]{0.99\linewidth}
 		\centering
 		\subfloat[player 1]
		{
 			\begin{overpic}[width=1\textwidth]
 	 			{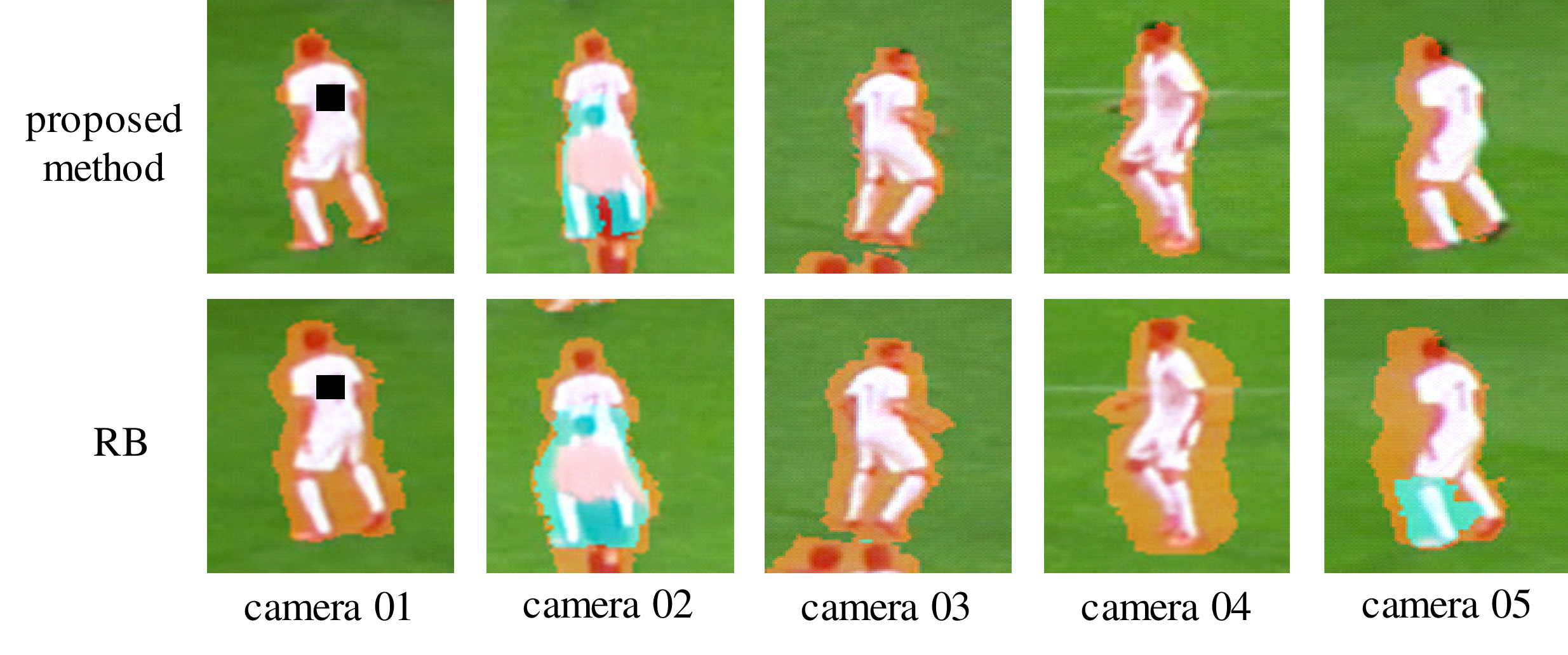}
 		\end{overpic}
 		}
	\end{minipage}	
\vskip 2mm
	\begin{minipage}[b]{0.99\linewidth}
 		\centering
 		\subfloat[player 2]
		{
 			\begin{overpic}[width=1\textwidth]
 	 			{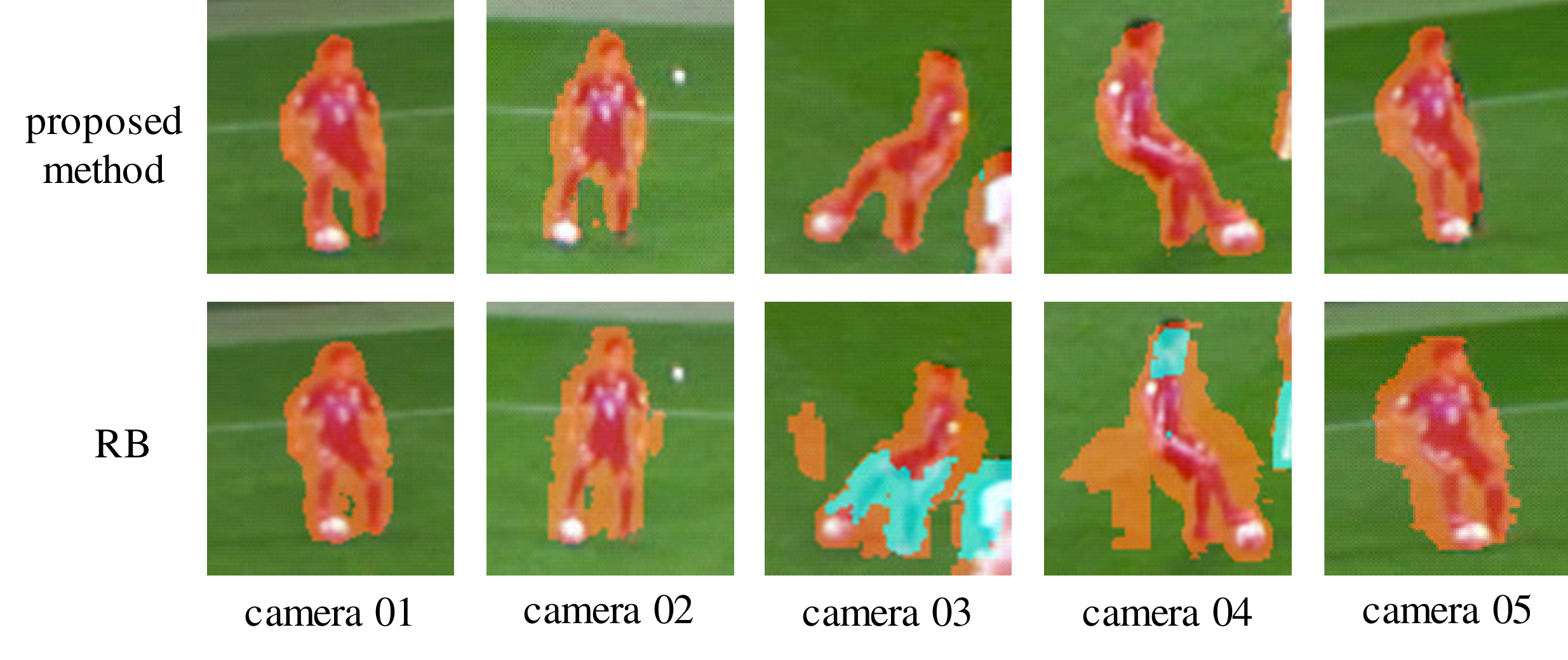}
 		\end{overpic}
 		}
	\end{minipage}	
\caption{Projections of billboard models of the first content on capturing viewpoints. The red and blue masking respectively indicate the visible and occluded regions.}
\label{fig:ffv-saudi-reprojection}
\end{figure} 

\begin{figure}[pt]
\centering
\footnotesize
	\begin{minipage}[b]{0.485\linewidth}
 		\centering
 		\subfloat[proposed method]
		{
 			\begin{overpic}[width=1\textwidth]
 	 			{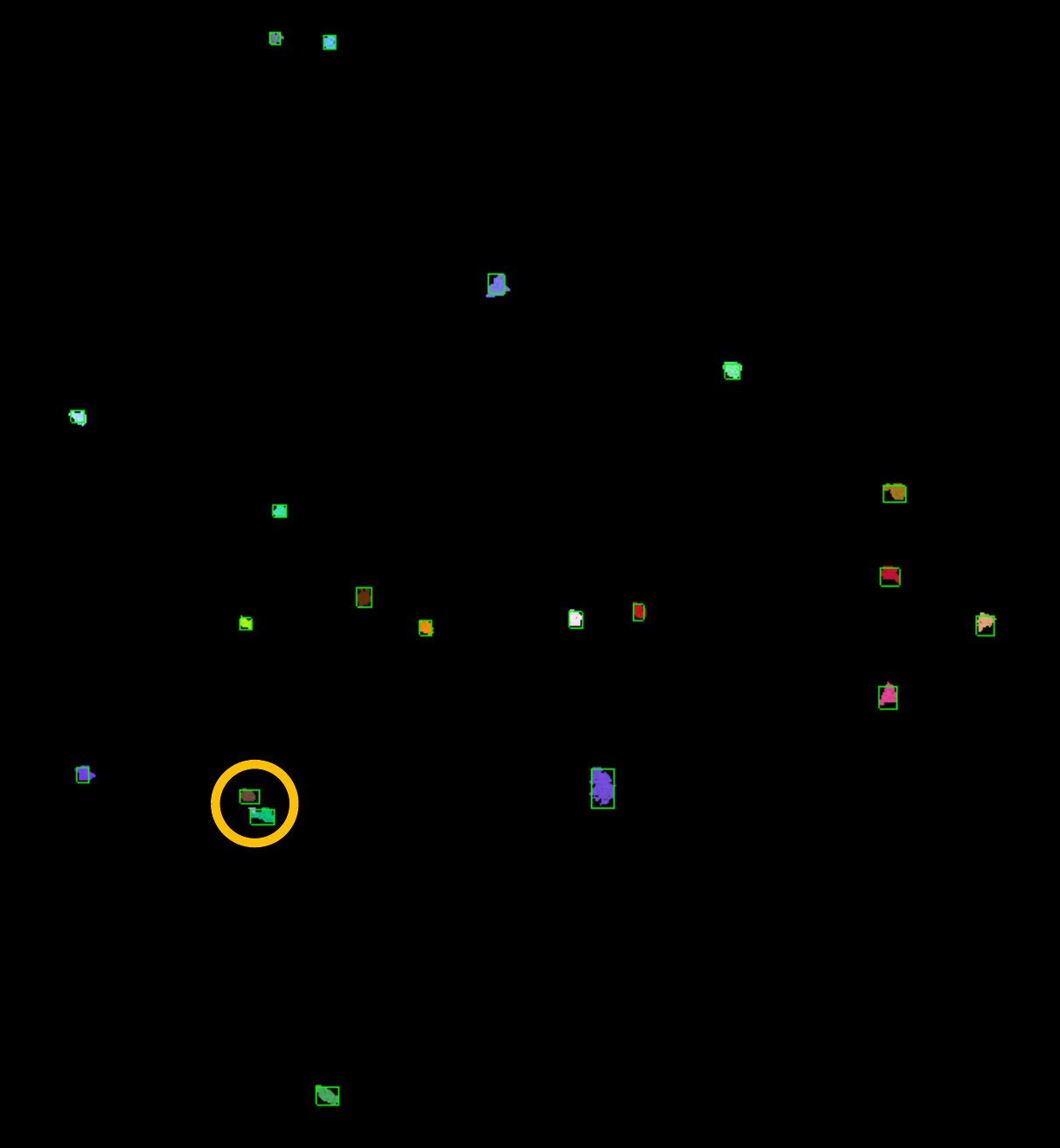}
 		\end{overpic}
 		}
	\end{minipage}	
	\begin{minipage}[b]{0.485\linewidth}
 		\centering
 		\subfloat[RB]
		{
 			\begin{overpic}[width=1\textwidth]
 	 			{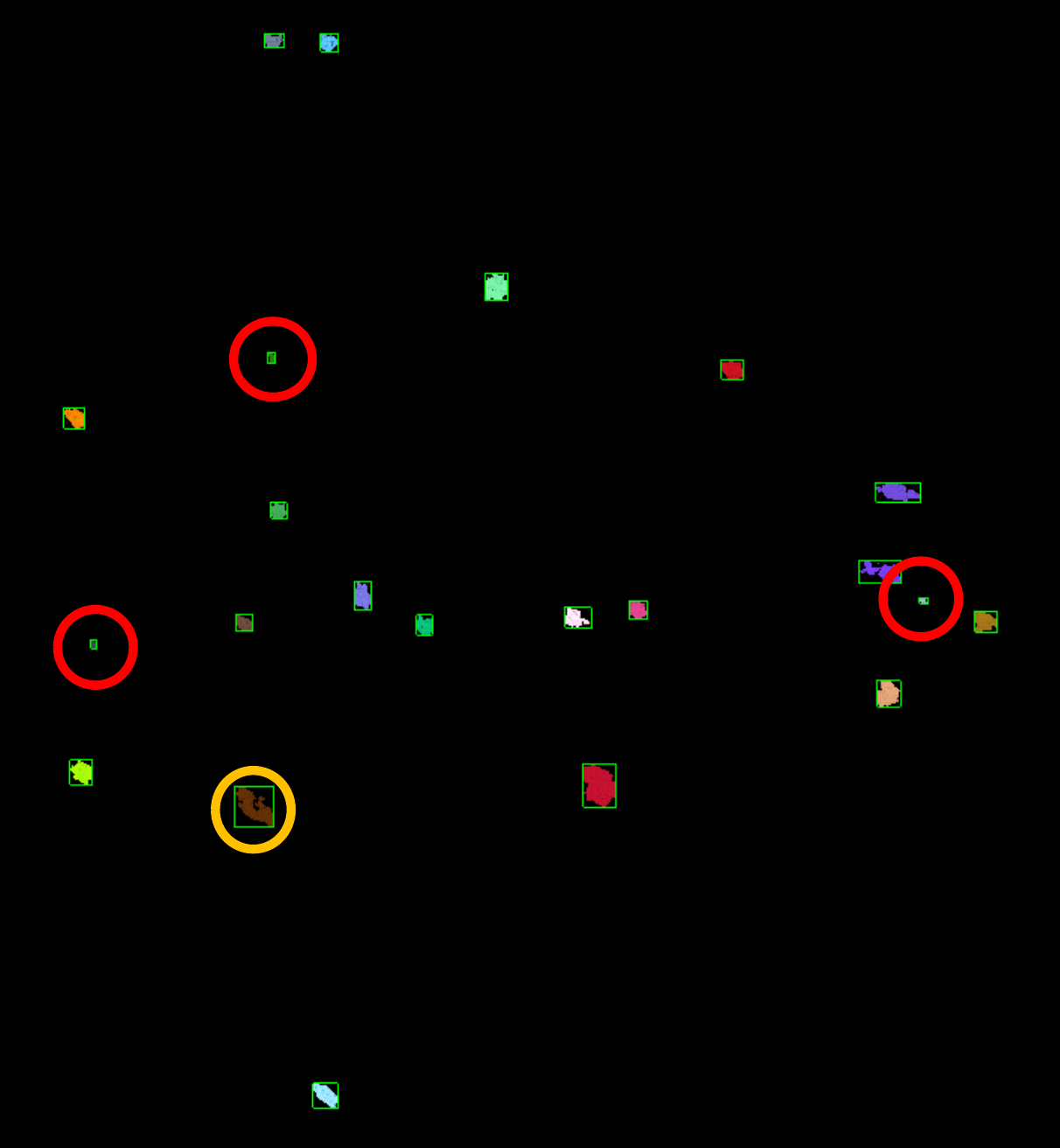}
 		\end{overpic}
 		}
	\end{minipage}	
\caption{Projections of the 3D model of the first content on the XY-plane. The red and yellow circles respectively highlight the noises and segmentation faults.}
\label{fig:ffv-saudi-xy-projection}
\end{figure} 

For the first content, we define the 3D space for reconstruction as $68$ meters wide, $4$ meters high, and $55.5$ meters deep.
The camera threshold of RB and CVH for the construction of a rough 3D shape was $4$, which remains the same in the production of the second content.
The voxel size for shape approximation in all the methods was $1$~cm $\times$ $1$~cm $\times$ $1$~cm.
{The thresholds, $T_{min}$ and $T_{max}$, for noise filtering were $3\times10^4$ and $3\times10^5$, respectively.}

Fig.~\ref{fig:ffv-saudi} (a) shows the cropped input image of each recording camera to highlight the region covered by virtual cameras.
Fig.~\ref{fig:ffv-saudi} (b) presents comparisons of three virtual viewpoint images produced by the proposed and reference methods, respectively.
Fig.~\ref{fig:ffv-saudi} (c) and (d) present the surface texture of two selected objects.
First, let us focus on the reproduced images from the first virtual viewpoint, shown in the first row of Fig.~\ref{fig:ffv-saudi} (b), (c), and (d).
The viewpoint was set with the same direction with ``cam01'' so that the methods (proposed method, RB, and FFVV) employing view-dependent rendering techniques can produce a high-quality texture. 
Nevertheless, CVH that utilizes global rendering techniques fails to give a proper appearance due to the inaccurate 3D shape approximation.
Next, let us look at the images constructed from the second virtual viewpoint, shown in the second row of Fig.~\ref{fig:ffv-saudi} (b), (c), and (d).
The virtual camera was set as bird's-eye from the above whose nearest reference camera is ``cam02''.
{It can be seen that the proposed method successfully recovers the color appearance of an occluded object.
However, the other techniques introduce severe artifacts or leave some important parts unrendered.}
Finally, let us observe the images (last row of Fig.~\ref{fig:ffv-saudi} (b), (c), and (d)) rendered by a virtual camera that was placed on the opposite side of ``cam01''.
Since the input images did not provide sufficient information for interpolation, the full model expression methods, FFVV and CVH, were incapable of offering a suitable chromatic appearance.  
However, the billboard methods have the potential to handle situations like this because they represent objects using planar billboard models that obtained from the nearest camera.

To illustrate the differences between our method and RB, we projected the billboard models back to the capturing viewpoints. 
The region mapped with a billboard model is marked with orange or blue. 
Orange means that the region is visible, while blue indicates an overlapped area.
Fig.~\ref{fig:ffv-saudi-reprojection} presents examples of projections of two objects.
{Comparison shows that the billboard models of our method are reliable and accurate, while RB tends to expand the individual object region and make a wrong judgment for occlusion.}
In the meantime, we projected the 3D models used in our method and RB onto the XY-plane to reveal the difference of 3D models, as shown in Fig.~\ref{fig:ffv-saudi-xy-projection}.
It can be seen that the model reconstructed by RB contains many noises that are highlighted by red circles in the figure.
Moreover, RB mistakenly recognizes two separate objects as one object, as demonstrated by the yellow circle in Fig.~\ref{fig:ffv-saudi-xy-projection}.

\begin{figure*}[pt]
\centering
\footnotesize
	\begin{minipage}[b]{0.95\linewidth}
 		\centering
 		\subfloat[{Cropped input images}]
		{
 			\begin{overpic}[width=1\textwidth]
 	 			{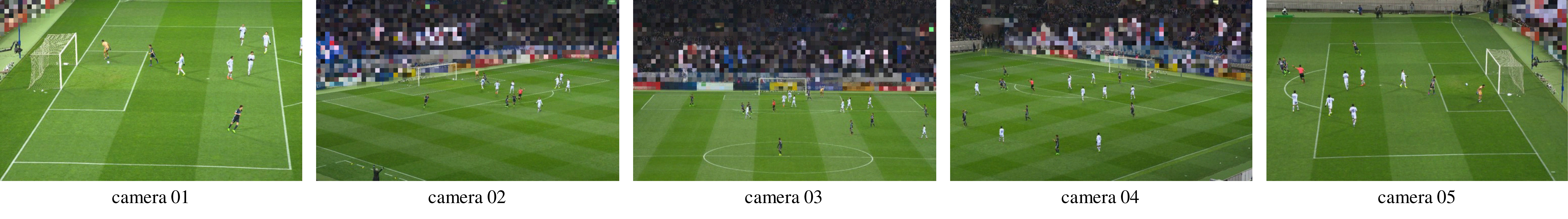}
 		\end{overpic}
 		}
	\end{minipage}	
\vskip 1mm

	\begin{minipage}[b]{0.95\linewidth}
 		\centering
 		\subfloat[Synthesized free-viewpoint video viewing from three virtual viewpoints]
		{
 			\begin{overpic}[width=1\textwidth]
 	 			{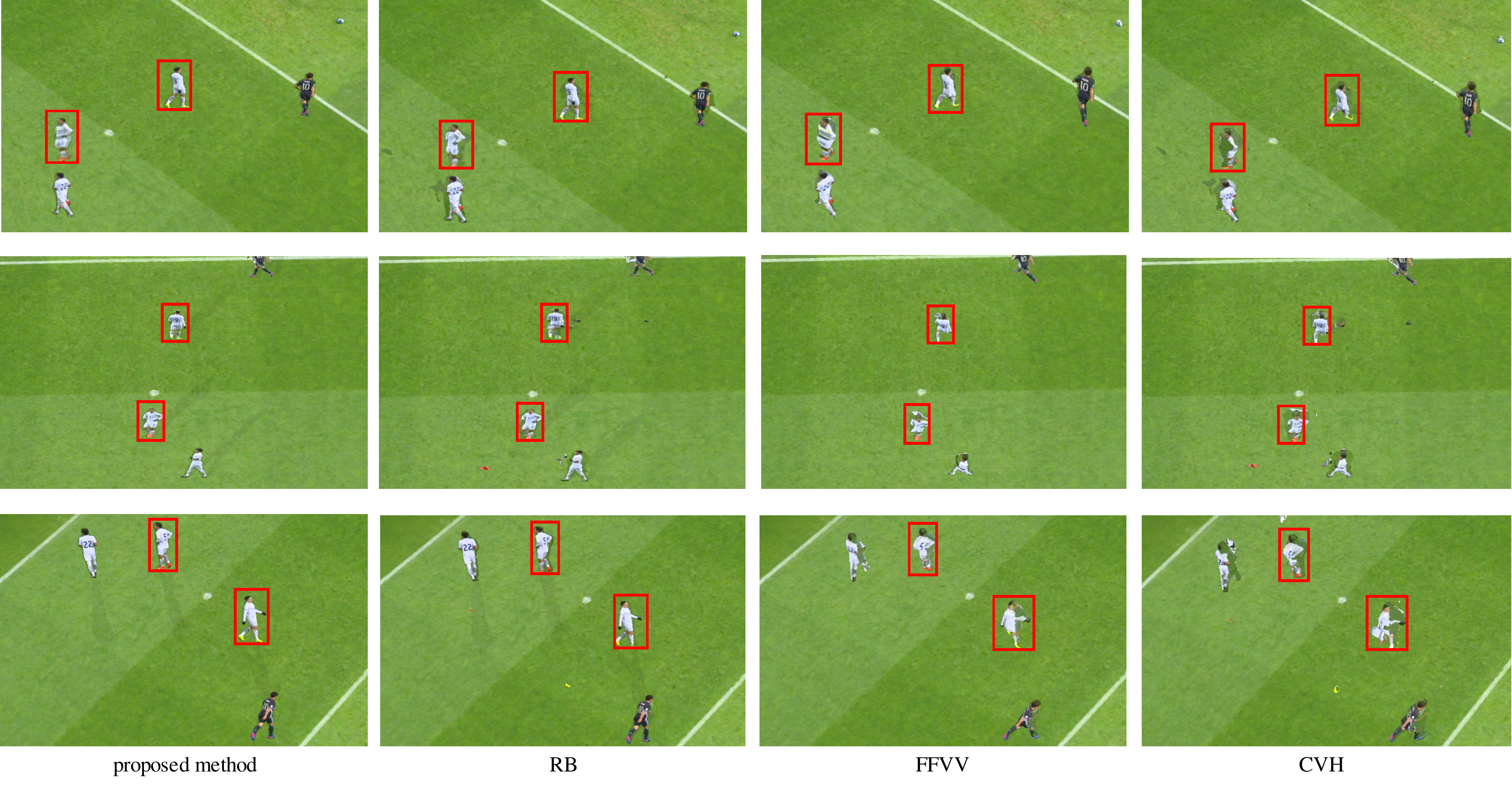}
 		\end{overpic}
 		}
	\end{minipage}	
\vskip 2mm
	\begin{minipage}[b]{0.463\linewidth}
 		\centering
 		\subfloat[Close up view of a selected player]
		{
 			\begin{overpic}[width=1\textwidth]
 	 			{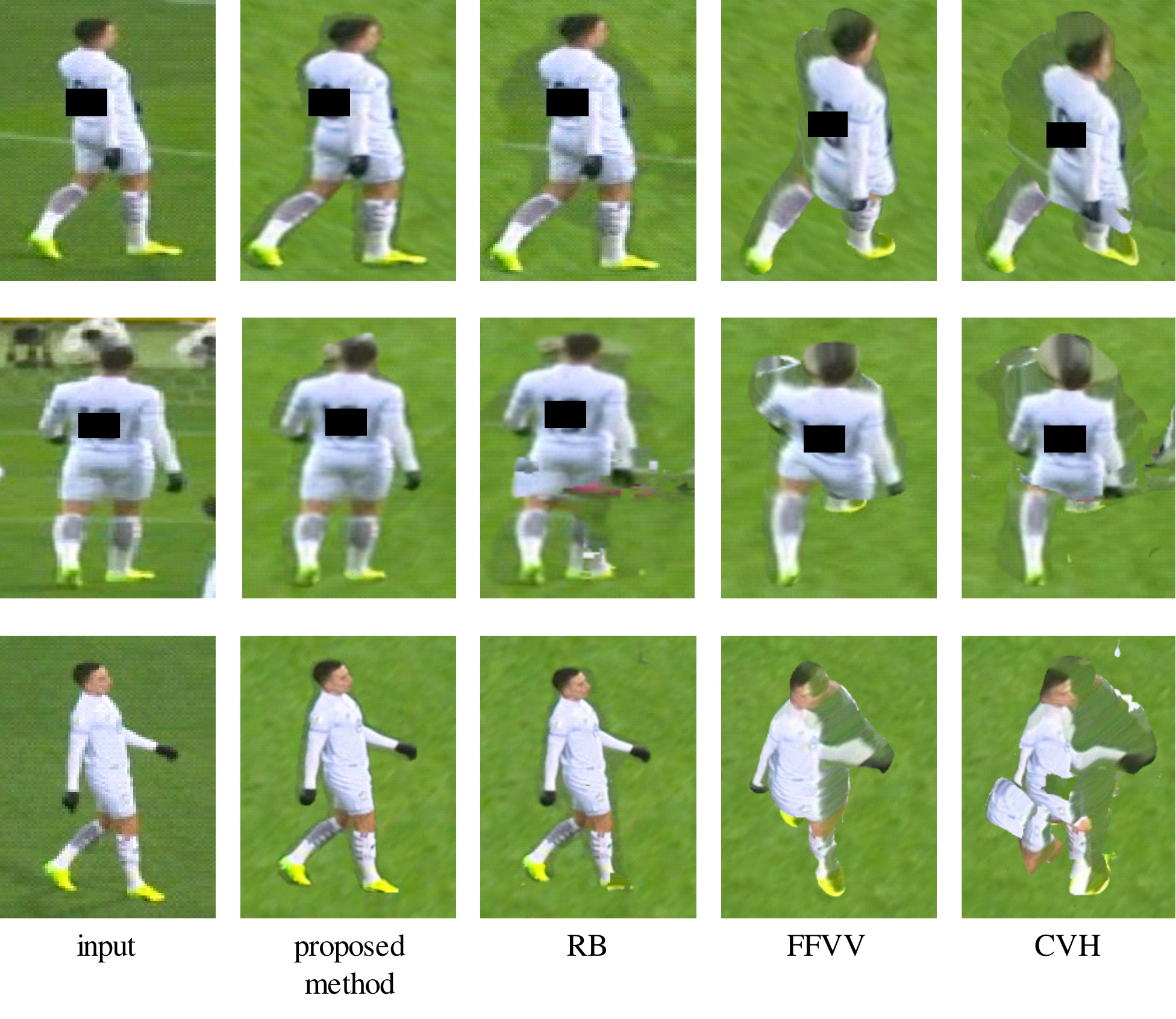}
 		\end{overpic}
 		}
	\end{minipage}	
\hskip 4mm
	\begin{minipage}[b]{0.465\linewidth}
 		\centering
 		\subfloat[Close up view of another selected player]
		{
 			\begin{overpic}[width=1\textwidth]
 	 			{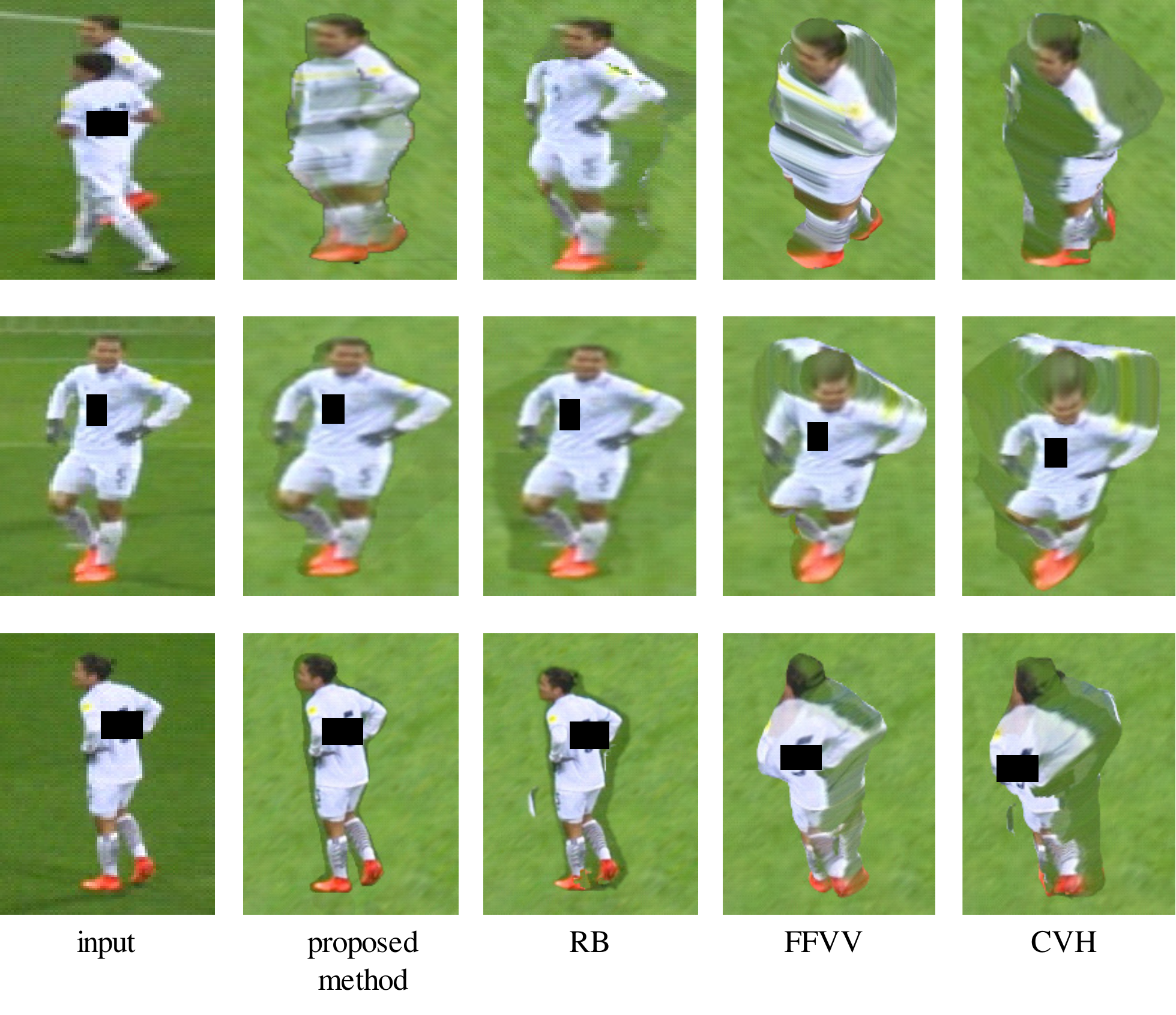}
 		\end{overpic}
 		}
	\end{minipage}	
\caption{Free-viewpoint video of the second content.}
\label{fig:ffv-thi}
\end{figure*} 

\begin{figure}[pt]
\centering
\footnotesize
	\begin{minipage}[b]{0.99\linewidth}
 		\centering
 		\subfloat[player 1]
		{
 			\begin{overpic}[width=1\textwidth]
 	 			{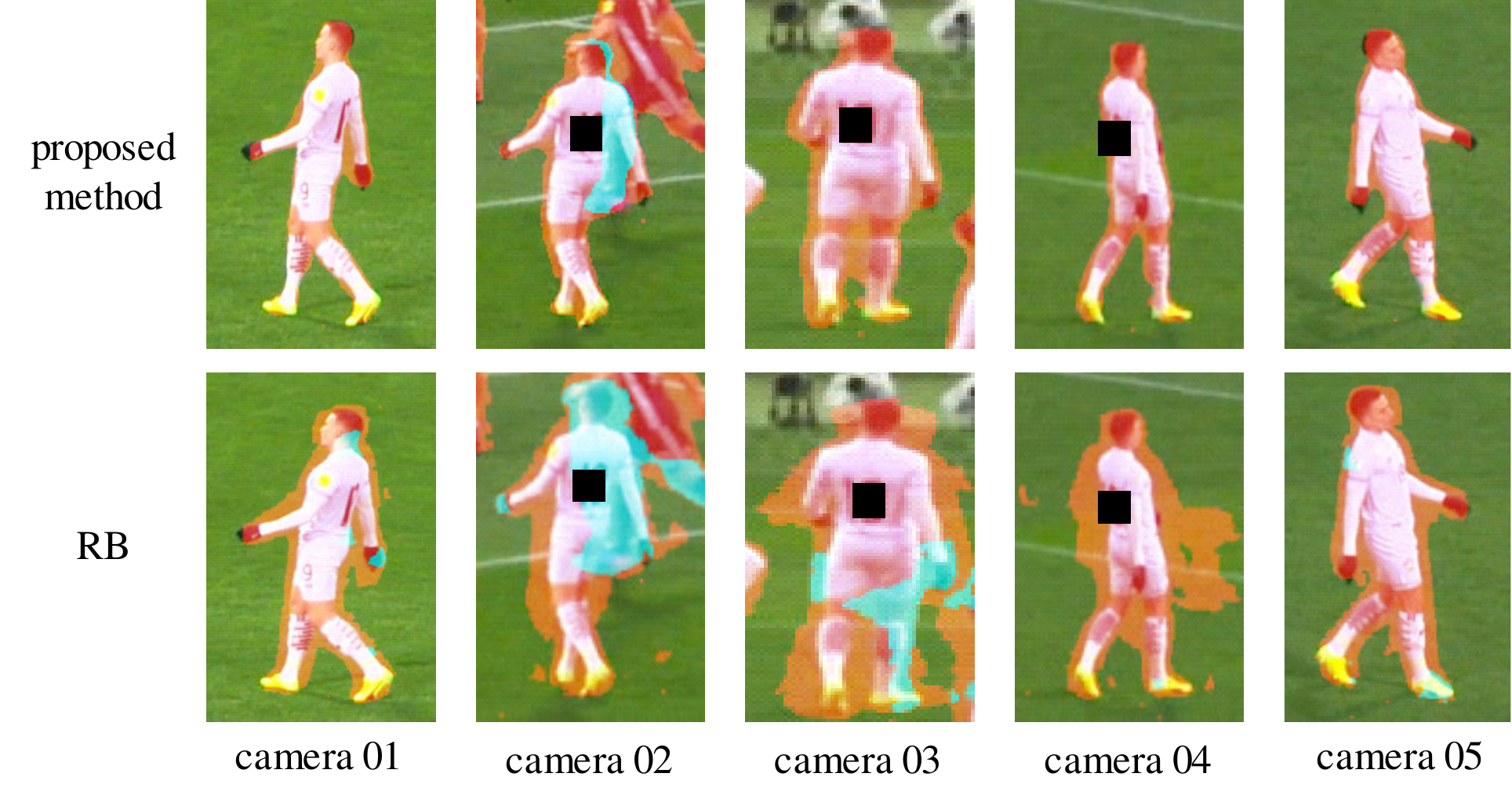}
 		\end{overpic}
 		}
	\end{minipage}	
\vskip 2mm
	\begin{minipage}[b]{0.99\linewidth}
 		\centering
 		\subfloat[player 2]
		{
 			\begin{overpic}[width=1\textwidth]
 	 			{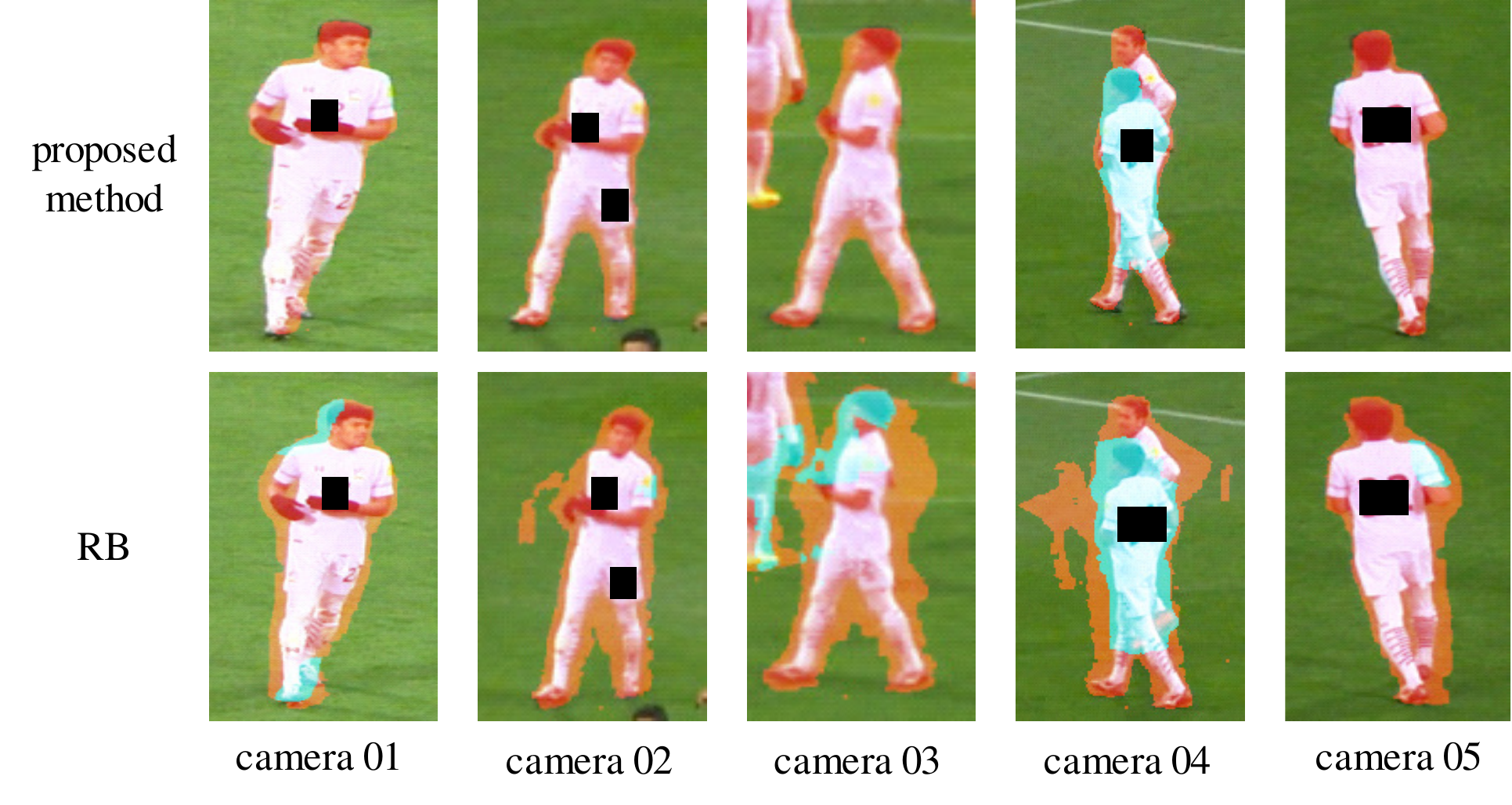}
 		\end{overpic}
 		}
	\end{minipage}	
\caption{Projections of billboard models of the second content on capturing viewpoints.}
\label{fig:ffv-thi-reprojection}
\end{figure} 

\begin{figure}[pt]
\centering
\footnotesize
	\begin{minipage}[b]{0.485\linewidth}
 		\centering
 		\subfloat[proposed method]
		{
 			\begin{overpic}[width=1\textwidth]
 	 			{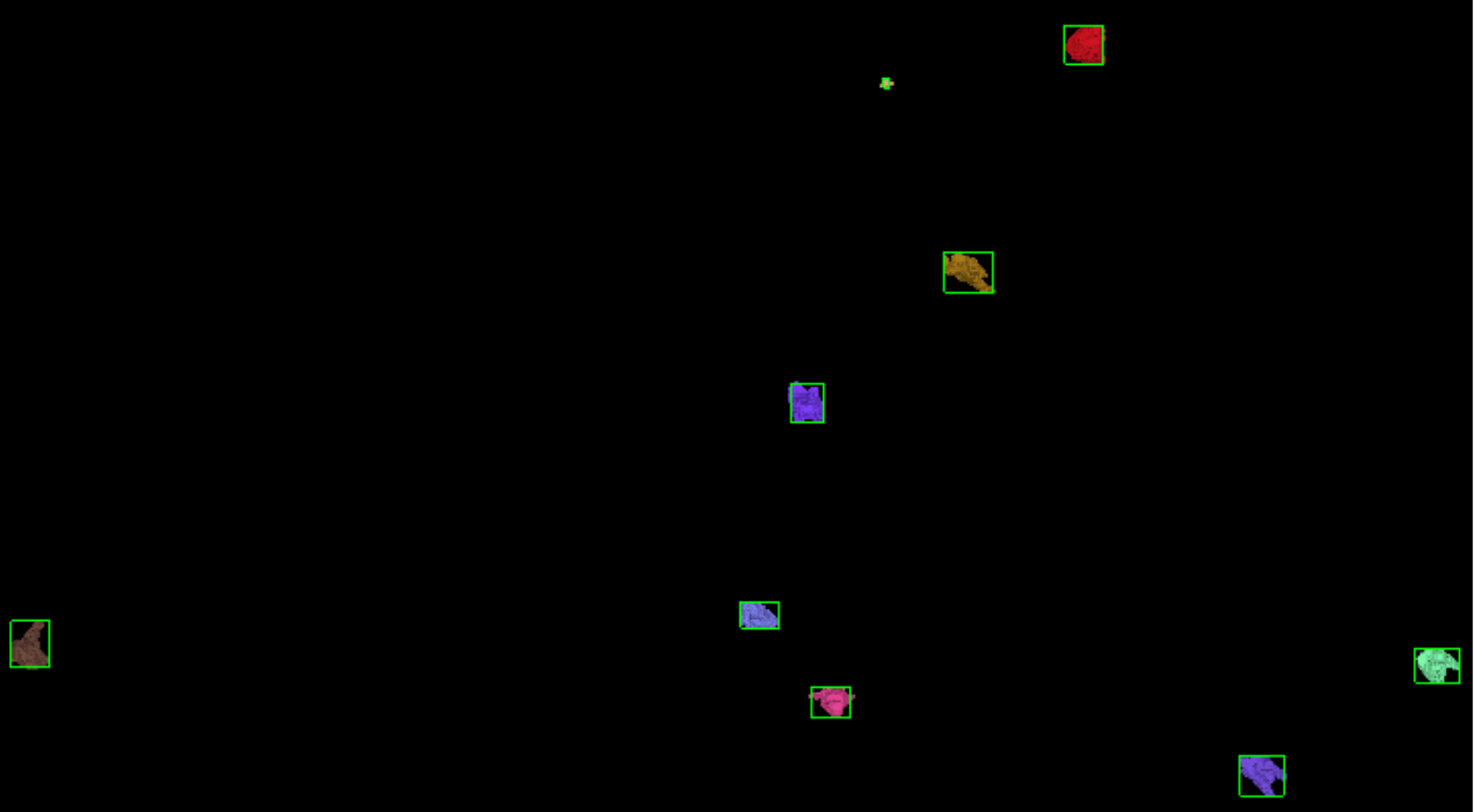}
 		\end{overpic}
 		}
	\end{minipage}	
	\begin{minipage}[b]{0.485\linewidth}
 		\centering
 		\subfloat[RB]
		{
 			\begin{overpic}[width=1\textwidth]
 	 			{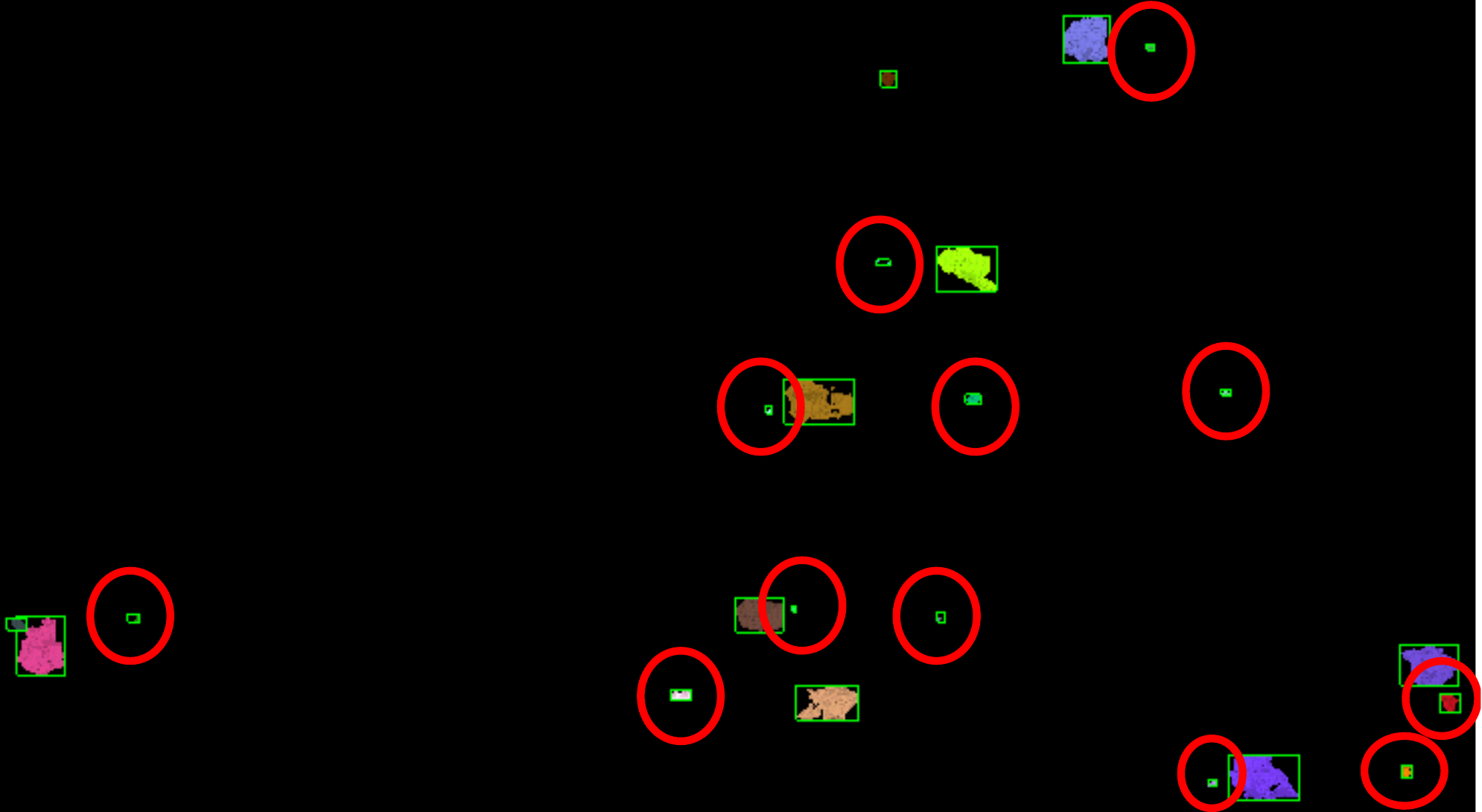}
 		\end{overpic}
 		}
	\end{minipage}	
\caption{Projections of 3D model of the second content on XY-plane.}
\label{fig:ffv-thi-xy-projection}
\end{figure} 

For the second content, the 3D space for production and the voxel size for shape approximation in all the methods were defined as $55$~m $\times$ $4$~m $\times$ $23$~m and $0.5$~cm $\times$ $0.5$~cm $\times$ $0.5$~cm, respectively.
{The thresholds, $T_{min}$ and $T_{max}$, for noise filtering were $2.4\times10^5$ and $2.4\times10^7$, respectively.}
Fig.~\ref{fig:ffv-thi} demonstrates the input images, synthesized images from three virtual viewpoints, and the highlighted surface texture of two selected objects.
All the virtual viewpoints were set as bird's-eye from above to evaluate the texture quality when the virtual facing directions are far from the recording directions.
Concerning the result in the figure, it can be observed that our method and RB, acting as billboard-based methods, outperforms the full model representation approach in all the tests.
The reason for this phenomenon is that the shape approximated from five cameras featuring wide-baseline is quite inaccurate.
The horizontal slice of a reconstructed model is more likely to be a pentagon but not a circle or ellipse with a smooth edge.
Thus the rendering quality is far from satisfactory.
Next, let us focus on the difference between the proposed method with RB.
{Besides the misalignment in rendering an occluded area, it can be seen that there are several artifacts or noise in the result of RB (the second row of Fig.~\ref{fig:ffv-thi} (c) and the third row of Fig.~\ref{fig:ffv-thi} (d)).}
The relaxed shape-from-silhouette approach is likely to introduce noises with irregular shape and size, as shown in Fig.~\ref{fig:ffv-thi-xy-projection}. 
Consequently, parts of the visible region in some cameras are judged to be occlusion, as demonstrated in Fig.~\ref{fig:ffv-thi-reprojection}.
Even though RB developed some noise filtering approaches, it is a challenging task to remove all noises, especially when their shapes resemble a ball.

\section{Discussion}

In this section, we discuss some factors that may affect the visual effect of a reconstructed free-viewpoint video.
First, camera calibration plays a vital role in free-viewpoint video creation.
Most of the reported approaches work based on the assumption that a sports field, such as a soccer field or rugby field, is the same as a design drawing.
However, the assumption fails in most cases.
This is sometimes because of human error when marking an actual sports field.
Moreover, sports associations usually provide rough guidelines, but not a specific number with reliable precision.
For example, soccer field dimensions are within the range found optimal by FIFA: $110-120$~yards ($100-110$~m) long by $70-80$~yards ($64-73$~m) wide.
Thus the camera calibration is not accurate enough, leading to errors in 3D shape reconstruction and texture rendering. 
Second, a camera-network should be laid out as carefully as possible to create a high-quality free-viewpoint video.
The primary requirement is that the cameras should be distributed uniformly in a stadium.
This setup is more likely to get well-rounded texture information that enhances the quality of reproduced surface appearances. 
In addition, this setup can provide continuous changes when switching viewpoint because a virtual view is represented by the billboard model of its nearest recording camera.
{A third factor is the number of cameras. 
There is no doubt that the more equipped cameras there are, the better.
However, we recommend the full model representation to be made as if there is an unlimited number of cameras.
The proposed method should receive top priority when only a small number of cameras is provided.
From our experience, five cameras are sufficient to create a high-fidelity free-viewpoint video.}
Finally, our method is appropriate for scenes involving many players, such as soccer, rugby, and basketball, but not suitable for simple scenarios with few players, such as judo, taekwondo, and wrestling.
The proposed method creates a stereo visual effect by placing 2D billboard model on different positions of a virtual stadium.
The scenarios with fewer players create fewer billboard models in each camera. Especially when the players grapple with each other, the proposed method only constructs one billboard model in each camera.
When all the players are represented by one billboard model, the spatial relationships among players are lost, making their 3D visual effectiveness weak.

\section{Conclusion}

In this paper, we presented a novel billboard-based synthesis approach suitable for free-viewpoint video production for sports scenes. It converts 2D images captured by a synchronized camera network to a high-quality 3D video.
Our approach has high flexibility because only a few cameras are required.
Therefore, it can apply to challenging shooting conditions where the cameras are sparsely placed around a wide area.
We approximate 3D models of objects using a conventional shape-from-silhouette technique and then project them onto each image plane to extract individual object regions and discover occlusions.
Each object region is rendered by a view-dependent approach in which the textures of non-occluded portions are taken from the nearest camera, while several cameras are used to reproduce the appearance of occlusions. 
Experimental results of soccer contents have proved that the surface texture of each object, including occluded ones, can be reproduced more naturally than by the other state-of-the-art methods.
In the future, we will parallelize our method and combine it with efficient data compression and streaming methods for delivering real-time free-viewpoint video.


%

\ifCLASSOPTIONcaptionsoff
  \newpage
\fi



%
%
%




\bibliographystyle{IEEEbib}
\bibliography{mybibfile.bib}

\end{document}